\documentclass[aps,prd,showpacs,twocolumn,superscriptaddress,nofootinbib,preprintnumbers]{revtex4-1} 

\makeatletter
\def\p@subsection{}
\makeatother

\usepackage{color,graphicx,amsmath,amssymb,lipsum}
\usepackage[colorlinks=true,citecolor=blue,linkcolor=blue,urlcolor=blue, backref=false,pdfborder={0 0 0}]{hyperref}

\usepackage[normalem]{ulem}
\usepackage[utf8]{inputenc} 
\usepackage{mathtools}
\usepackage{wrapfig}
\usepackage{hyperref}

\usepackage{float}

\usepackage{ulem}
\usepackage{diagbox}

\usepackage[dvipsnames]{xcolor}

\newcommand{\be}{\begin{equation}}
\newcommand{\ee}{\end{equation}}
\newcommand{\beqa}{\begin{eqnarray}}
\newcommand{\eeqa}{\end{eqnarray}}

\renewcommand\L{\Lambda}

\newcommand{\kmin}{k_{\rm min}}
\newcommand{\kmax}{k_{\rm max}}
\newcommand{\hMpc}{\,h \text{Mpc}^{-1}}

\newcommand{\eff}{{\rm eff}}

\newcommand{\kNL}{k_{\rm NL}}

\newcommand{\eV}{\,\text{eV}}

\newcommand{\Gpc}{\,\text{Gpc}}
\newcommand\Mpc{\text{Mpc}}

\newcommand{\bseq}{\begin{subequations}}
\newcommand{\eseq}{\end{subequations}}

\renewcommand{\ln}{\mathop{\rm ln}\nolimits}

\def\gsim{\raise0.3ex\hbox{$\;>$\kern-0.75em\raise-1.1ex\hbox{$\sim\;$}}}
\def\lsim{\raise0.3ex\hbox{$\;<$\kern-0.75em\raise-1.1ex\hbox{$\sim\;$}}}

\def\beqn#1{\begin{equation}\label{#1}}
\def\eeqn{\end{equation}}

\def\beqa#1{\begin{eqnarray}\label{#1}}
\def\eeqa{\end{eqnarray}}

\def\Z2{$\mathcal{Z_2}$}

\newcommand {\ignore}[1]{}

\begin{document}


\title{Cosmological constraints from the power spectrum of eBOSS quasars}

\author{Chudaykin Anton}\email{Chudayka@mcmaster.ca}\affiliation{Department of Physics \& Astronomy, McMaster University,\\ 
1280 Main Street West, Hamilton, ON L8S 4M1, Canada}
\affiliation{Perimeter Institute for Theoretical Physics, Waterloo,
  Ontario, N2L 2Y5, Canada} 

\author{Mikhail M.~Ivanov}\email{ivanov@ias.edu}\affiliation{School of Natural Sciences, Institute for Advanced Study, 1 Einstein Drive, Princeton, NJ 08540, USA}
\affiliation{NASA Hubble Fellowship Program Einstein Postdoctoral Fellow}

\begin{abstract} 
We present 
the effective-field theory (EFT)-based 
cosmological full-shape analysis of
the anisotropic power spectrum of eBOSS quasars at the effective redshift $z_{\rm eff}=1.48$. 
We perform extensive tests of our pipeline on simulations, 
paying a particular attention to the modeling of observational systematics, such as 
redshift smearing, fiber collisions, and the radial integral constraint.
Assuming the minimal $\Lambda$CDM model, and fixing the primordial power spectrum tilt
and the physical baryon density, we find the Hubble constant $H_0=(66.7\pm 3.2)~$km~s$^{-1}$Mpc$^{-1}$, the matter density fraction $\Omega_m=0.32\pm 0.03$, and the 
late-time mass fluctuation amplitude $\sigma_8=0.95\pm 0.08$.
These measurements
are fully consistent with the Planck cosmic microwave background results. 
Our eBOSS quasar $S_8$ posterior, $0.98\pm0.11$,
does not exhibit 
the so-called $S_8$ tension.
Our work paves the way for systematic full-shape analyses of quasar samples
from future surveys like DESI. 
\end{abstract}

\maketitle

\section{Introduction}

The distribution of luminous objects on large
cosmological scales (large-scale structure) 
is one of the key observables 
that allows us to understand the expansion 
history and constituents of our Universe. 
Large-scale structure traces Universe's evolution 
at low redshifts that are especially important for 
dark energy studies. As such, it provides important
information complementary to that of the cosmic microwave 
background radiation (CMB) measurements (e.g. {\it Planck}~\cite{Planck:2018vyg}, ACT~\cite{ACT:2020gnv}, SPT-3G~\cite{SPT-3G:2021eoc}), whose primary fluctuations are most sensitive
to physics at a much earlier epoch, during recombination.

The role of various large-scale structure 
probes in measuring cosmological parameters
has become even more prominent in the recent 
years due to the appearance of  
tensions between various cosmological datasets. 
The most critical ones are the Hubble and growth tensions, which manifest themselves in the difference between direct and indirect 
probes of the Hubble constant $H_0$ and the structure formation parameter $S_8$ (or equivalently,
the 
mass fluctuation amplitude $\sigma_8$), see~\cite{Abdalla:2022yfr} for a recent review.
Large scale structure surveys such as BOSS~\cite{BOSS:2016wmc}, eBOSS~\cite{eBOSS:2020yzd}, DES~\cite{DES:2017myr,DES:2021wwk}, KIDS~\cite{KiDS:2020suj}, HSC~\cite{HSC:2018mrq}, 
are already contributing very significantly in efforts to 
understand these tensions.
In particular, the extended Baryon acoustic Oscillation Spectroscopic Survey (eBOSS)
has provided crucial probes of the expansion rate through 
baryon acoustic oscillations (BAO) and the matter growth through redshift-space
distortions (RSD)~\cite{eBOSS:2020yzd}.

An important development in spectroscopic survey data analysis 
has appeared recently thanks to theoretical efforts in the effective field theory (EFT)
of large-scale structure~\cite{Baumann:2010tm,Carrasco:2012cv}, see~\cite{Cabass:2022avo}
for a review. This theory gives an accurate and mathematically 
consistent theoretical model for clustering of matter and various luminous tracers 
in the mildly-nonlinear (quasilinear) regime. 
This regime is relevant for shot-noise limited spectroscopic surveys,
which contain the bulk of cosmological information precisely on the quasilinear scales characterized by wavenumber $k\lesssim 0.5~\hMpc$. It should be 
emphasized that the EFT is systematic and consistent, i.e. 
it is a program of successive approximations
that allows us to compute the nonlinear clustering of galaxies to 
any desired accuracy. This can be contrasted with popular  
phenomenological models, whose regime of validity, and hence accuracy, is ultimately 
limited even on large scales.

Importantly, the EFT model allows to describe the entire shape of the galaxy 
power spectrum, and hence extract the information which is not accessible 
with the conventional BAO/RSD techniques~\cite{Ivanov:2019pdj,DAmico:2019fhj}. 
The EFT-based full-shape analysis has been successfully applied to the 
power spectra and bispectra of BOSS luminous red galaxies (LRGs)~\cite{Ivanov:2019pdj,DAmico:2019fhj,Philcox:2020vvt,Philcox:2021kcw,Chen:2021wdi} and also the 
power spectrum of eBOSS emission line 
galaxies (ELGs)~\cite{Ivanov:2021zmi}. 
The main goal of our paper is to extend the EFT-based full-shape
analysis to the eBOSS quasars (QSO)~\cite{Ata:2017dya,Hou:2018yny,Hou:2020rse,Neveux:2020voa}.

Quasars are interesting for multiple reasons. Thanks to their high luminosity, 
they can be used to trace the 
matter distribution at high redshifts. 
In particular, the eBOSS quasar sample covers the redshift range $0.8<z<2.2$.
This gives us a new window onto the physics of our Universe just before the onset 
of dark energy. In addition, the large comoving volume covered by quasars 
makes them ideal to constrain local primordial non-Gaussianity~\cite{Castorina:2019wmr,Mueller:2021tqa}, 
which, if present, should manifest itself as a scale-dependent bias 
on very large scales~\cite{Desjacques:2016bnm}. 
These are the reasons that made eBOSS QSO a subject of intense research 
over last years~\cite{Leistedt:2014zqa,Ata:2017dya,Hou:2018yny,Castorina:2019wmr,Hou:2020rse,Neveux:2020voa,Mueller:2021tqa,Semenaite:2021aen,Neveux:2022tuk}. 
Application of the EFT framework to the QSO promises 
to multiply the scientific gain from this sample,
which can be estimated by a successful execution of the EFT program on 
the eBOSS LRG and ELG catalogs~\cite{Philcox:2021kcw,Ivanov:2021zmi}.
In this work we make a step towards systematic 
studies of the eBOSS QSO sample with the EFT of LSS. 

Our paper is structured as follows. We present a summary of main results 
in Section~\ref{sec:sum}. We describe the data and our method in details in Section~\ref{sec:data}. Extensive tests on simulations and mock catalogs 
are presented in Section~\ref{sec:mock}.
Our final results are presented in Section~\ref{sec:res}
and discussed in Section~\ref{sec:disc}. 
Several appendices contain important additional material, 
such as the full parameter and constraint tables, 
further tests of systematics, and prior volume effects.

\section{Summary of main results}
\label{sec:sum}

We start with a short summary of our results. We have analysed the publicly available redshift-space power spectrum monopole ($\ell=0$), quadrupole ($\ell=2$) and hexadecopole ($\ell=4$) moments of the eBOSS quasar sample~\cite{Neveux:2020voa}. 
This sample covers the redshift range $0.8<z<2.2$ (effective redshift $z_{\rm eff}=1.48$)
and is spreads across two different patches of the sky: North Galactic Cap (NGC) and South Galactic Cap (SGC). 
In our baseline QSO analysis we fix the current physical baryon density $\omega_b=0.02268$ using the BBN prediction \cite{Aver:2015iza,Cooke:2017cwo}, and adopt the primordial power spectrum tilt $n_s=0.9649$ from the Planck CMB measurements \cite{Planck:2018vyg}~\footnote{We checked that imposing tight Gaussian priors on $\omega_b$ and $n_s$ from BBN and Planck, respectively, leads to identical parameter constraints.}.
We vary three other cosmological parameters in our Markov Chain Monte Carlo (MCMC) chains: the physical density of cold dark matter ($\omega_{cdm}$), the reduced Hubble constant ($h$) and the logarithm of the primordial scalar power spectrum amplitude, $\ln(10^{10}A_s)$. We approximate the neutrino sector with a single state of mass $m_\nu=0.06\eV$ and two massless states. Alongside cosmological parameters, we vary a set of $26$ nuisance parameters ($13$ per each cap) capturing non-linear clustering of matter, galaxy bias, and redshift-space distortions.

\begin{figure*}[ht!]
\begin{minipage}{0.5\textwidth}
\centering
\includegraphics[width=0.95\textwidth]{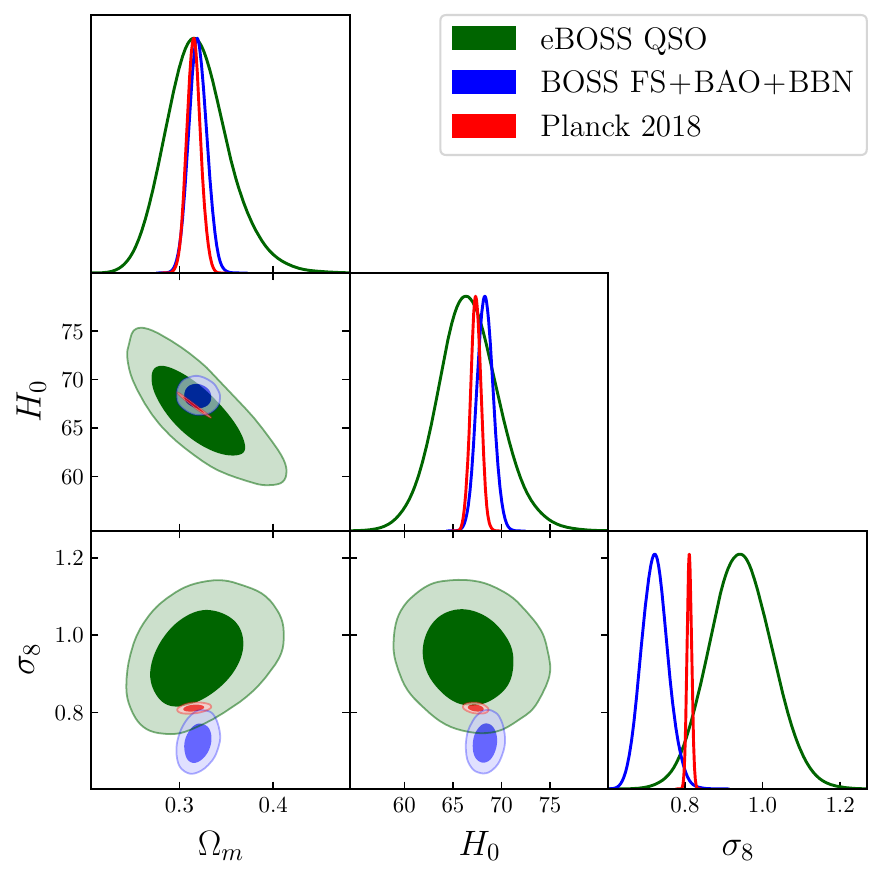}
\end{minipage}%
\centering
 \begin{minipage}{0.5\textwidth}
\includegraphics[width=0.95\textwidth]{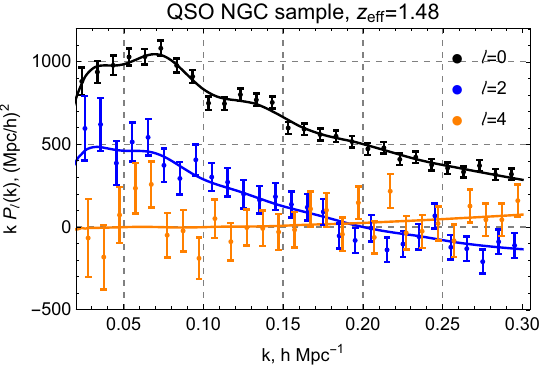}
\includegraphics[width=0.95\textwidth]{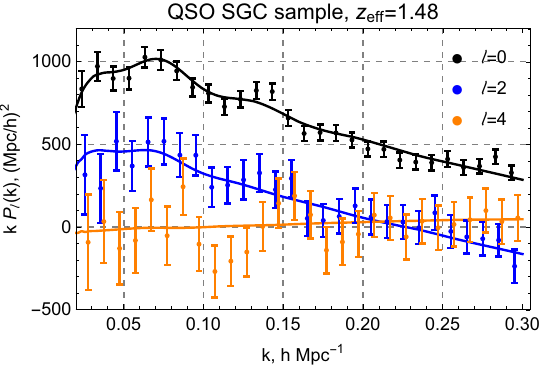}
\end{minipage}
\caption{
\textit{Left panel:} marginalized constraints on the matter fraction $\Omega_m$, Hubble constant $H_0$ and mass fluctuation amplitude $\sigma_8$ in the $\Lambda$CDM model from the power spectra of eBOSS quasars (QSO). For comparison we also show the Planck CMB 2018 \cite{Planck:2018vyg} and BOSS DR12 full-shape analysis of the galaxy power spectra and 
the bispectrum monopole \cite{Philcox:2021kcw}. \textit{Right panel:} QSO power spectrum monopole ($\ell=0$), quadrupole ($\ell=2$) and hexadecapole ($\ell=4$) moments from North and South Galactic Caps of the eBOSS data. The best-fit $\Lambda$CDM model from the combined analysis of both caps is shown by solid lines. The error bars correspond to the square root of diagonal elements of the covariance matrix.
\label{fig:small} } 
\end{figure*}

The posterior distribution of the QSO eBOSS data in $\Omega_m-H_0-\sigma_8$ space 
is shown in the left panel of Fig.~\ref{fig:small} along with the BOSS full-shape analysis of the red luminous galaxy sample and the Planck 2018 results. The right panel of Fig. \ref{fig:small} shows the QSO $P_\ell(k)$ measurements along with the best-fit theoretical model. The 1d marginalized constraints are listed in Tab.~\ref{tab:main}.

\begin{table*}[t!]
  \begin{tabular}{|c||c|c|c|} \hline
   \diagbox{ {\small Param.}}{\small Dataset}  
   & eBOSS QSO 
   & BOSS FS+BAO+BBN
   &  \textit{Planck} 2018
      \\ [0.2cm]
\hline
$\omega_{cdm}$   
& $0.1184_{-0.011}^{+0.009}$
& $0.1262_{-0.0059}^{+0.0053}$
& $0.1200_{-0.0012}^{+0.0012}$
\\ 
\hline
$10^2\omega_{b}$   
& $-$
& $-$
& $2.237_{-0.015}^{+0.015}$
\\ 
\hline 
$h$   
& $0.6669_{-0.034}^{+0.031}$
& $0.6832_{-0.0086}^{+0.0083}$
& $0.6736_{-0.0053}^{+0.0055}$
  \\ \hline
$\ln(10^{10}A_s)$  
& $3.375_{-0.18}^{+0.18}$
& $2.742_{-0.099}^{+0.096}$
& $3.044_{-0.015}^{+0.014}$
\\ 
\hline
$n_s$  
& $-$
& $-$
& $0.9649_{-0.0042}^{+0.0042}$
\\ 
\hline
\hline
$\Omega_m$   
& $0.3205_{-0.038}^{+0.03}$
& $0.3196_{-0.01}^{+0.01}$
& $0.3153_{-0.0077}^{+0.0071}$
\\ \hline
$\sigma_8$   
& $0.9445_{-0.083}^{+0.081}$
& $0.7221_{-0.037}^{+0.032}$ 
& $0.8112_{-0.0062}^{+0.006}$ 
\\ \hline
$S_8$   
& $0.976_{-0.12}^{+0.1}$
& $0.745_{-0.041}^{+0.037}$ 
& $0.832_{-0.013}^{+0.013}$ 
\\ \hline
\end{tabular}
\caption{
Mean values and 68\% CL minimum credible
intervals for the parameters of the $\L$CDM model.
The BBN prior on $\omega_b$ and the Planck prior on $n_s$ are assumed in the two LSS analyses, and the corresponding posteriors are not displayed.
The top group are the parameters directly sampled in the MCMC chains, the bottom ones
are derived parameters.}
\label{tab:main}
\end{table*}

\section{Data and methodology}
\label{sec:data}

\subsection{Data}
\label{sec:data1}

{\bf QSO.} Our main analysis is based on the power spectra from the QSO SDSS DR16 catalogue \cite{Neveux:2020voa}. The QSO sample was selected from the SDSS-I-II-III optical imaging data in the {\it ugriz} photometric pass bands \cite{Fukugita:1996qt} and from the Wide-field Infrared Survey Explorer \cite{Wright:2010qw}. For details of the selection algorithm see Ref. \cite{BOSS:2015wkd}. The eBOSS QSO power spectra were estimated by drawing the redshifts of the random catalogues from the data catalogue, which suppresses the radial modes on large scales. We refer the reader to Refs. \cite{Neveux:2020voa,deMattia:2019vdg} for details on the QSO catalogue and its systematics.

The power spectrum was 
built from the galaxies
collected 
from two different patches of the sky. 
The NGC and SGC patches have the following effective redshifts $z_\eff$, effective shot noise term $P_0^{\rm shot}$ of the Yamamoto estimator \cite{Yamamoto:2005dz,Neveux:2020voa}~\footnote{Note that the shot noise term includes various corrections for survey geometry and systematic effects. Thus, it does not identically correspond to a physical number density of the sample.}
and comoving geometric volumes $V$, 
\be 
\begin{split}
 \text{NGC}:~&z_{\rm eff}=1.48\,,~ P_0^{\rm shot}=6.3\cdot 10^4~[\text{Mpc}/h]^3\,,\\
&  V=13.62~h^{-3}\text{Gpc}^3\,,
\end{split}
\ee
\be 
\begin{split}
\text{SGC}:~&z_{\rm eff}=1.48\,,~ P_0^{\rm shot}=7.1\cdot 10^4~[\text{Mpc}/h]^3\,,\\
&  V=8.76~h^{-3}\text{Gpc}^3\,.
\end{split}
\ee

In our analysis we use the publicly available pre-reconstructed eBOSS QSO power spectrum miltipoles computed from the FKP-weighted density field \cite{Feldman:1993ky} using the Yamamoto estimator \cite{Yamamoto:2005dz} implemented in the \texttt{nbodykit} package \cite{Hand:2017pqn}.\footnote{Publicly available at \url{https://svn.sdss.org/public/data/eboss/DR16cosmo/tags/v1_0_1/dataveccov/lrg_elg_qso/QSO_Pk/}} 
We use the monopole, quadrupole and hexadecope moments with the scale cuts 
$\kmin=0.02$ and $\kmax=0.3\hMpc$ following the eBOSS analysis \cite{Neveux:2020voa} and our own tests presented later. The power spectrum multipoles are binned in momentum space spheres of width $\Delta k=0.01 \hMpc$. The redshifts and angular scales are converted into comoving coordinates using a flat $\Lambda$CDM cosmology with $\Omega_m=0.31$.

{\bf BOSS FS+BAO+BBN.} We also present the results of the full-shape BOSS DR12 LRG analysis based on the power spectrum and the bispectrum monopole data from \cite{BOSS:2016wmc,Gil-Marin:2016wya} combined with the BBN prior on $\omega_b$ and the Planck prior on $n_s$. We use the same likelihood as in Ref.~\cite{Philcox:2021kcw} based on the window-free estimators.~\footnote{Publicly available at \url{https://github.com/oliverphilcox/BOSS-Without-Windows}}
The power spectrum analysis includes the monopole, quadrupole and hexadecope moments with $\kmin=0.01$ and $\kmax=0.2\hMpc$ along with the BAO measurements performed with the post-reconstructed power spectra. We compute the cross-covariance between the BAO and full-shape measurements from the Patchy mocks according to the procedure presented in Ref. \cite{Philcox:2020vvt}. As far as the bispectrum is concerned we use $\kmin=0.01$ and $\kmax=0.08\hMpc$ with step $\Delta k=0.01$ which corresponds to 62 unique triangular configurations following the recent analyses \cite{Ivanov:2021kcd,Philcox:2021kcw}.

\subsection{Theory model}
\label{sec:data2}

We fit the QSO power spectrum multipoles' data using the one-loop perturbation theory (effective field theory). The theoretical predictions are calculated using the \texttt{CLASS-PT} code \cite{Chudaykin:2020aoj}. 
This code is based on the path 
integral formulation of the EFT of LSS
known as time-sliced perturbation 
theory~\cite{Blas:2015qsi,Blas:2016sfa,Ivanov:2018gjr,Ivanov:2018lcg,Vasudevan:2019ewf}. 
In this work we vary the following set of power spectrum nuisance parameters in our MCMC chains, 
\be
\{b_1,b_2,b_{\mathcal{G}_2},b_{\Gamma_3},c_0,c_2,c_4,\tilde{c},a_2,P_{\rm shot},P_{\rm shot,2}\}\,,
\ee
where 
in addition to the standard one-loop parameters \cite{Ivanov:2019pdj,Chudaykin:2020ghx} we introduce a constant contribution to the quadrupole moment, $P_{\rm shot,2}$, which serves to model systematic effects present in the QSO sample. We will validate our model with N-body and approximate simulations. The priors on these nuisance parameters will be discussed shortly. Detailed information about the standard EFT of LSS theoretical model
and nuisance parameters 
can be found in Refs.~\cite{Ivanov:2019pdj,Chudaykin:2020aoj}.

\subsection{Window function}
\label{sec:data3}

\begin{figure}[!thb]
\centering
\includegraphics[width=0.49\textwidth]{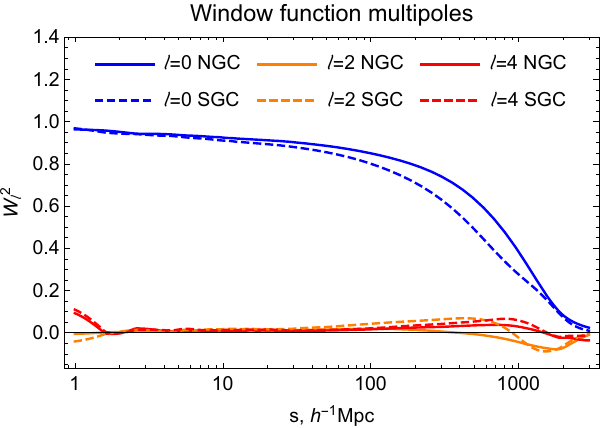}
\caption{Window function multipoles of the baseline eBOSS QSO sample.
\label{fig:win} } 
\end{figure}

{\bf Survey geometry.} To take survey geometry effects into account one needs to fold the true power spectrum with a survey-specific selection function. We use the publicly available survey window functions from \cite{Neveux:2020voa}. The eBOSS QSO window function multipoles for the different galactic caps are shown in Fig. \ref{fig:win}.

\begin{figure}[!t]
\centering
\includegraphics[width=0.49\textwidth]{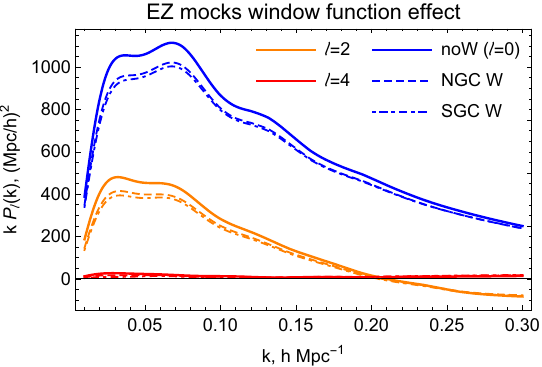}
\caption{Effect of the window functions on typical power spectrum multipoles of the EZ mocks for NGC and SGC footprints.
\label{fig:win2} } 
\end{figure}

We implement window function by employing the standard approach based on the plane parallel approximation, see Ref. \cite{Wilson:2015lup,BOSS:2016psr}. This method amounts to a simple multiplication of the true correlation functions $\xi^{\rm true}_\ell$ in position space with the window function multipoles $W_\ell^2$. In our analysis we use the quasar multipole moments up to the hexadecopole, so we have 
\be
\begin{split}
& \xi^{\rm win}_0=\xi^{\rm true}_0 W_0^2 
+ \frac{1}{5}\xi^{\rm true}_2 W_2^2+\frac{1}{9}\xi^{\rm true}_4 W_4^2\,,\\
& \xi^{\rm win}_2=\xi^{\rm true}_0 W_2^2 + 
\xi^{\rm true}_2\left[
W_0^2 + \frac{2}{7}W_2^2 + \frac{2}{7}W_4^2 
\right] \\
&\quad \quad \quad+ \xi^{\rm true}_4\left[
\frac{2}{7} W_2^2 + \frac{100}{693}W_4^2 
\right] \,,\\
& \xi^{\rm win}_4=\xi^{\rm true}_0 W_4^2 + 
\xi^{\rm true}_2\left[
\frac{18}{35}W_2^2 + \frac{20}{77}W_4^2 
\right]\\
& \quad \quad \quad + 
\xi^{\rm true}_4\left[W_0^2+
\frac{20}{77}W_2^2 + \frac{162}{1001}W_4^2 
\right]\,.
\end{split} 
\ee
We have checked that the effect of higher order window functions is negligible. Given this reason, we do not include multipole window functions with $\ell=6$ and higher. 
The effect of the window functions on the power
spectrum multipoles is shown in Fig. \ref{fig:win2}.

\begin{figure}[!thb]
\centering
\includegraphics[width=0.49\textwidth]{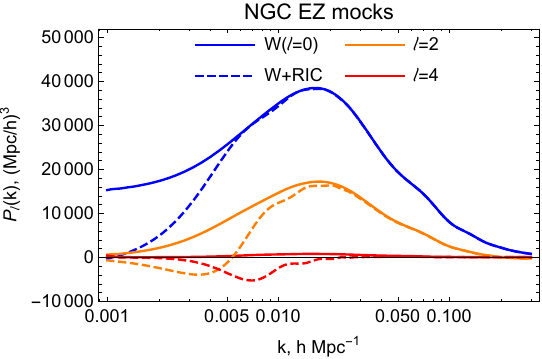}
\caption{The impact of the 
integral constraint on typical power spectrum multipoles of the EZ mocks for the NGC footprint.
\label{fig:w_ric} }
\end{figure}

{\bf Radial integral constraint.} 
Another important effect is due to the scheme used to assign redshifts to randoms from the mock data redshift distribution which suppresses radial modes on large scales. 
To take this effect into account, we implement the radial integral constraint (RIC) using the formalism introduced in Ref. \cite{deMattia:2019vdg}. The full RIC correction is to be subtracted from the windowed correlation function in configuration space,
\be
\label{eq:ic}
\begin{split}
\xi_\ell^{\rm icc}(s)&=\xi_\ell^{\rm win}(s)-P_{\rm shot}\mathcal{W}^{\rm sn}_{\ell}(s)\\
&-\sum_{\ell'}
\frac{4\pi}{2\ell'+1}\int s'^2ds'~\mathcal{W}_{\ell \ell'}(s,s')
\xi_{\ell'}^{\rm true}(s')\,,
\end{split} 
\ee
where $\mathcal{W}_{\ell \ell'}$ are Legendre multipole moments of the 3-point correlator of the selection function, and $\mathcal{W}^{\rm sn}_{\ell}$ denotes the shot noise contribution to the integral constraint, for details we refer the reader to Ref. \cite{deMattia:2019vdg}. Following this work, 
we treat separately 
the deterministic piece 
of the power spectrum 
and the term produced by the 
constant shot noise contribution. 
This procedure increases numerical accuracy. 
Thus, the correlation functions $\xi^{\rm true}_\ell$ in Eq.~\eqref{eq:ic}
does not contain the shot noise contribution, i.e. one needs to perform Fourier transforms of power spectrum {\it without} the shot noise correction $P_0^{\rm shot}$.
Note that according to Ref. \cite{Neveux:2020voa}, the impact of the RIC constraint on cosmological inference is below one percent for the scale cut $\kmin=0.02~\hMpc$.
Nevertheless, keeping in mind future analyses of primordial non-Gaussianity sensitive to large
scales, 
we prefer to account for the RIC in the theory model by implementing Eq.~\eqref{eq:ic}. 
The effect of the RIC for the NGC footprint is shown in Fig. \ref{fig:w_ric}.

{\bf Fiber collisions.} Finally, we implement the correction due to fiber collisions. 
We follow the effective window
approach proposed in \cite{Hahn:2016kiy}. In this method, two colliding targets are modeled by a top-hat window function that depends on the physical scale of fiber collisions, $D_{\rm fc}$. In Fourier space the effect of fiber collisions on the true power spectrum is captured by two terms: Fourier transform of the top-hat function (uncorrelated part) and the power spectrum convolved with the top-hat filter (correlated part). The first contribution reads~\cite{Hahn:2016kiy}
\be
\label{uncorr}
\Delta P_\ell^{\rm uncorr}(k)=-f_s(2\ell+1)L_\ell(0)\frac{(\pi D_{\rm fc})^2}{k}W_{2D}(kD_{\rm fc})\,,
\ee
where $L_\ell$ are Legendre polynomials, $W_{2D}(x)=2J_1(x)/x$ is the cylindrical top-hat function in 2D, and $f_s$ is the fraction of the survey affected by fibre collisions. 
The correlated part is given by \cite{Hahn:2016kiy}
\be
\label{corr}
\Delta P_\ell^{\rm corr}(k)=-f_sD_{\rm fc}^2\int\frac{d^2q_{\perp}}{(2\pi)^2}P(k_\parallel,q_\perp)W_{2D}(q_\perp D_{\rm fc})\,,
\ee
where $k_\parallel$ is the projection along the line of sight, and $q_\perp$ denote 2D modes perpendicular to the line of sight. Strictly speaking, this integral receives unreliable contributions which must be renormalized by a set of polynomials in $k$ \cite{Hahn:2016kiy}. However, the QSO window functions are nearly scale independent on large scales (see Fig. \ref{fig:win}), so unreliable contributions for monopole and quadrupole would exactly coincide,
at leading order, with the EFT stochastic couterterms $P_{\rm shot}$ and $a_2$ that are already marginalized over in our analysis. For the hexadecopole the unreliable contribution scales as $k^4$ and hence represents a higher order stochastic term keeping which is beyond the accuracy 
of our one-loop model. 
Given these reasons, we do not introduce free parameters to correct computations of the 
fiber collision convolution integrals.

For the QSO sample the collision radius is $D_{\rm fc}=0.9\hMpc$ \cite{Neveux:2020voa}. This is the comoving distance that corresponds to the fiber collision angular scale $62''$ at the effective redshift of the QSO sample $z_\eff=1.48$. The fractions of the survey affected by fiber collisions for two different footprints read \cite{Neveux:2020voa}
\be
\label{fig:fc}
f_s(\text{NGC})=0.36\,,\quad f_s(\text{SGC}) =0.45\,.
\ee
We perform the integration of the power spectrum in eq. \eqref{corr} over $q\in[5\cdot10^{-4},0.4]\hMpc$. We checked that the final cosmological constraints are stable against the choice of
the integration limits once the appropriate counterterms are added to the model.

In Fig. \ref{fig:fc} we illustrate the impact of the effective window approach to account for the systematic effects applied to the NGC EZ mock catalogs.
\begin{figure}[!t]
\centering
\includegraphics[width=0.49\textwidth]{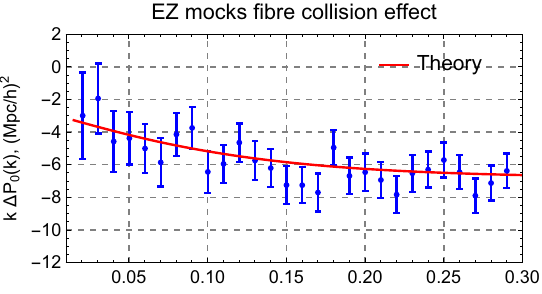}
\includegraphics[width=0.49\textwidth]{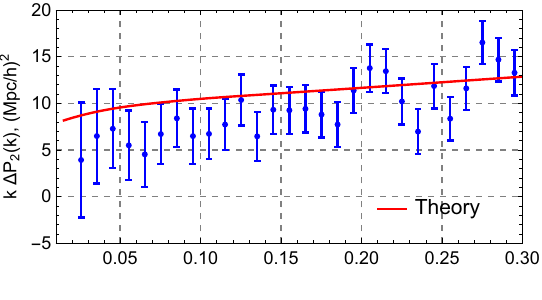}
\includegraphics[width=0.49\textwidth]{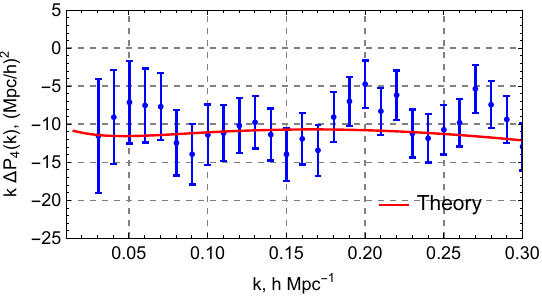}
\caption{Shifts on the means of power spectrum multipoles induced by the systematic effects applied to the NGC EZ mocks (blue dots). The red line shows the theoretical prediction for the fibre collision effect given by the sum of \eqref{uncorr} and \eqref{corr}. Error bars corresponds to the variance estimated from 1000 realizations of the EZ mocks. 
\label{fig:fc} } 
\end{figure}
The total size of the fiber collision correction modeled by \eqref{uncorr} and \eqref{corr} is shown as a red line that is in good agreement with the observed systematic shifts in the NGC EZ mocks for all multipoles. Note that the blue points show the change in the power spectra when all systematic effects are applied.
It implies that the largest systematic offset arises from fiber collisions, in agreement with conclusions of Ref.~\cite{Neveux:2020voa}. 
Notice that in 
Fig.~\ref{fig:fc} we 
display only the deterministic parts 
of the fiber collision corrections. The residual offset between the data 
and the theory curves in that figure
is accounted for by the counterterms.
Once the nuisance parameters are 
fitted from the data, 
the agreement between the 
data and the theory is restored.

A similar level of agreement is found for the SGC footprint.

\subsection{Redshift uncertainties}
\label{sec:smear}

Measurements of quasar clustering is affected by redshift uncertainties that strongly impact  parameter recovery. The leading effect is the redshift smearing due to broad emission lines in the quasar spectra. This effect originates due to fast rotation of gas in the black holes. This gas is affected by winds which offset the emission lines in quasars' spectra. 
The quasar clustering is also affected by catastrophic redshifts, where the redshift estimate happens to be very far from the true value due to line confusion or contamination in the sky. Around $1.5\%$ of objects in the eBOSS QSO sample have been assigned catastrophic redshifts \cite{Smith:2020stf}. 
Finally, there are redshift uncertainties due to spectroscopic measurements. Reliable modeling of all those effects is essential in order to obtain unbiased cosmological measurements.

Previous works~\cite{Hou:2018yny,Hou:2020rse,Neveux:2020voa,Smith:2020stf} have proposed
to model redshift uncertainties with modified exponential fingers-of-God (FoG) prefactors. 
Practically, they use a damping term that comprises both a Gaussian and Lorentzian functions 
in order to model both the Gaussian and the non-Gaussian parts of the small-scale velocity distribution~\cite{BOSS:2016off}.
This FoG prefactor should be present in the phenomenological models of Refs.~\cite{Hou:2018yny,Hou:2020rse,Neveux:2020voa,Smith:2020stf} even in the absence of observational
systematics in order to account for the physical effect
of non-linear redshift space distortions. Refs.~\cite{Neveux:2020voa,Smith:2020stf}
have found that the standard FoG model successfully absorbs the redshift uncertainties due to redshift smearing effects as well. 

In the EFT context, the fingers of God effect is treated perturbatively 
by introducing higher derivative operators along the line of sight.  
One may hope that the standard EFT expansion would also absorb 
redshift uncertainties just as the standard
FoG model did so in the previous analyses.
We have found that this logic does not work completely. Indeed, 
we will show later that the redshift smearing
introduce a constant offset in the quadrupole moment on large scales. This offset 
cannot be fit by a $k^2 \mu^2 P_{\rm lin}$ or $k^4\mu^4 P_{\rm lin}$ counterterm 
($\mu$ is the line-of-sight angle). It cannot be fit by the standard
FoG model either.
To model this systematic effects present in the QSO sample we extend our EFT-based theoretical framework with a new constant term $P_{\rm shot,2}$ contributing to the quadrupole moment, 
\be
P_2(k)\rightarrow P_2(k)+P_{\rm shot,2}
\ee
Note that this term is forbidden in 
EFT of LSS~\cite{Chudaykin:2020aoj}, so it can only be 
generated by non-local observation systematics. We justify the inclusion of the $P_{\rm shot,2}$ in Sec. \ref{sec:mock1}.
We compare our EFT-based full-shape analysis with the standard treatment of systematic effects by means of the standard FoG damping term in App. \ref{app:FoG}.

\subsection{Parameter and priors}
\label{sec:data4}

Let us now discuss the choice of parameters and corresponding priors. We impose the following flat priors on $\Lambda$CDM parameters,
\be
\label{eq:priorcosm}
\begin{split}
& h\in \text{flat}[0.4,1]\,,\quad 
\omega_{cdm}\in \text{flat}[0.03,0.7]\,,\\
&\ln(10^{10}A_s)\in \text{flat}[0.1,10]
\end{split} 
\ee
The eBOSS quasar data cannot constrain the current physical baryon density $\omega_b$ and tilt of the primordial scalar power spectrum $n_s$ with a reasonable accuracy. For this reason, we 
fix 
\be
\omega_b=0.02268
\ee
from the BBN data \cite{Aver:2015iza,Cooke:2017cwo}. We also assume
\be
n_s=0.9649
\ee
from the Planck CMB measurements \cite{Planck:2018vyg}.
We approximate the neutrino sector with one single state of mass $m_\nu=0.06\eV$ and two massless states.

As far as bias parameters are concerned we use~\cite{Chudaykin:2020aoj},
\be
\label{biasPr}
\begin{split}
 b_1\in \text{flat}[0,4]\,, \quad &b_2\sim \mathcal{N}(0,1^2)\,, \\
 b_{\mathcal{G}_2}\sim \mathcal{N}(-0.4,1^2) \,, \quad
&b_{{\Gamma}_3}\sim \mathcal{N}\left(0.77,1^2\right) \,,
\end{split}
\ee
where the means in the second line are given by the co-evolution model \cite{Desjacques:2016bnm} with the linear bias $b_1=2.4$ from the official eBOSS QSO analysis \cite{Neveux:2020voa}. For higher derivative terms (counterterms) we assume
\be
\label{cssPr}
\begin{split}
& \frac{c_0}{[\text{Mpc}/h]^2} \sim \mathcal{N}(20,20^2)\,,\quad 
\frac{c_2}{[\text{Mpc}/h]^2} \sim \mathcal{N}(30,20^2)\,,\\
& \frac{c_4}{[\text{Mpc}/h]^2} \sim \mathcal{N}(20,20^2)\,,\quad 
\frac{\tilde{c}}{[\text{Mpc}/h]^4} \sim \mathcal{N}(0,200^2)\,,
\end{split}
\ee
which are motived by the EFT naturalness arguments in the physical units~\cite{Chudaykin:2020aoj}.
For the stochastic contributions we adopt
\be
\label{eq:shotpr}
\begin{split}
P_{\rm shot} \sim \mathcal{N}(0,1^2)\,,
\quad a_2\sim \mathcal{N}(0,1^2)\,,
\end{split}
\ee
Note that in our conventions the stochastic contributions are given by
\be
\label{eq:Pstoch}
P_{\rm stoch}=\frac{1}{\bar n}\left[1+
P_{\rm shot} 
+ a_2\mu^2 \left(
\frac{k}{k_{\rm NL}}\right)^2 
\right]\,,
\ee
with $\kNL=0.45\hMpc$. 
For $P_{\rm shot,2}$ which serves to model systematic effects present in the QSO sample we assume the following Gaussian prior,
\be
\label{pshot2}
P_{\rm shot,2}=-1000\pm1000\,(\Mpc/h)^3
\ee
We validate the choice of this prior with the accurate Outer Rim simulations and approximate light-cone EZ mocks.

Let us comment on the choice of priors w.r.t. previous analyses \cite{Chudaykin:2020aoj,Chudaykin:2020ghx}. First, we use the prior on $b_{\mathcal{G}_2}$ centered on the co-evolution model prediction. This choice is in agreement with the analysis of the accurate Outer Rim simulations which predict negative values of this parameter. 
Second, we impose $30\%$ tighter priors on the leading order counterterm coefficients and nearly two times stronger bound on $\tilde c$ compared to that considered in \cite{Chudaykin:2020aoj,Chudaykin:2020ghx}.
We have checked that using wider priors as in 
these references, 
 enhances the marginalization projection effects slightly, see App. \ref{app:ORnosys} for more detail.
Our priors for Finger-of-God counterterms are validated in the mock data analysis presented in Sec. \ref{sec:mock1}. The third point is that we do not include stochastic contribution proportional to $k^2$ into the theoretical model. This choice is motivated by field level 
analyses of high resolution N-body simulations~\cite{Schmittfull:2018yuk,Schmittfull:2020trd} that do not find evidence for this term for a sample similar to ours. 
We also found that including this contribution does not affect parameter recovery 
other than through increasing prior volume effects.
For these reasons we do not model this stochastic bias and set it to zero.  

In total we vary 3 cosmological parameters along with 13 nuisance parameters for NGC and SGC samples separately (26 in total).

\subsection{Likelihood}
\label{sec:data5}

We build up the Gaussian likelihood for power spectrum multipoles,
\be\begin{split}
& -2\ln \mathcal L_P = 
\sum_{i,j}(\tilde C^{\ell \ell'}_{ij})^{-1}
\Delta P_{\ell}(k_j)
\Delta P_{\ell'}(k_j)\,,\quad \text{where} \\
& \Delta P_{\ell}(k_i)\equiv 
P^{\rm theory}_{\ell}(k_j)- P^{\rm data}_{\ell}(k_j)
\,,
\end{split}
\ee
where $\ell,\ell'=0,2,4$. We use the publicly available covariance matrix estimated from 1000 realization of EZ mocks for each Galactic cap (NGC and SGC) \cite{Zhao:2020bib}. We correct it for the finite number of mocks following \cite{Hartlap:2006kj}:
\be
\label{Hart}
\tilde C^{-1}=\frac{N_{\rm mocks}-N_{\rm data}-2}{N_{\rm mocks}-1}C^{-1}\,,
\ee
where $N_{\rm mocks}$ is the total number of EZ mocks and $N_{\rm data}$ is the number of data points. Note that the EZ mocks have been generated with a higher shot-noise level than that in the real data that results in overestimation the monopole contribution by $3\%$ \cite{Neveux:2020voa}. 

Finally, let us note that, strictly speaking, if the covariance matrix
is estimated from a finite number of simulations, 
the likelihood 
needs to be adjusted to match a multivariate modified t-distribution \cite{Sellentin:2015waz}. 
We have explicitly
adjusted our MCMC code 
to sample this t-distribution
instead of the Gaussian likelihood, 
and have found nearly 
indistinguishable results. 
This agrees with previous 
tests
performed in Ref.~\cite{Philcox:2020zyp}.

\section{Tests on simulations}
\label{sec:mock}

We will test our pipeline now on two kinds of mocks. First, we will fit the Outer Rim mocks, 
which are based on exact N-body dynamics, and also contain accurately 
injected redshift smearing effects. The second kind of the mocks we will study are
EZ mocks, which are based on an approximate gravity solver, but include 
all necessary 
observational systematics, such as the window function effects and fiber collisions.

\subsection{Outer Rim mocks}
\label{sec:mock1}

To validate our pipeline and in particular the priors on nuisance parameter we study the Outer Rim (OR) QSO mocks \cite{Smith:2020stf}. These mocks were designed to reproduce statistical properties of eBOSS QSO sample and were used to validate eBOSS analysis pipeline \cite{Neveux:2020voa,Hou:2020rse}.
The catalogs were constructed from the Outer Rim N-body simulation at $z=1.433$ with volume $27\,h^{-3}\Gpc^3$. This simulation was run in a flat $\Lambda$CDM cosmology with the following fiducial parameters, 
\be
\begin{split}
& h=0.71\,,\quad \omega_{cdm}=0.1109\,,\quad  \omega_b = 0.02258\,,\\
& n_s= 0.963\,,\quad \sigma_8 = 0.8\,, \quad M_{\rm tot}=0~\text{eV}\,.
\end{split} 
\ee
Dark matter halos are populated with quasars using 20 halo occupation distribution (HOD) models. For each of the 20 HOD models, 100 independent random realizations have been created. Strictly speaking, the realizations for each HOD model are not independent because they are built from the same N-body simulation. 
But since the quasar duty cycle is low, the individual realizations are uncorrelated within the statistical precision of the mocks~\cite{Smith:2020stf}. 
The scatter between the mocks corresponding to the different HOD models is much smaller than the error in the QSO data which is estimated from EZ lightcone mocks~\cite{Zhao:2020bib}. Therefore, without loss of generality, we focus on the mock catalog generated with the HOD-3 model, which was picked as a reference sample in Ref.~\cite{Smith:2020stf}.

The Outer Rim QSO mocks have been used to derive the systematic uncertainties originating from the modeling of the QSO power spectra~\cite{Neveux:2020voa}. The main effect is from the redshift uncertainties due to broad emission lines in the quasar spectra which impacts the recovery of cosmological parameters.
To reliably model this effect, four different sets of mock have been created: without systematic effects, with Gaussian smearing, with non-Gaussian smearing and with non-Gaussian smearing+catastrophic redshifts. We refer the reader to Ref.~\cite{Smith:2020stf} for the details on the modeling of various systematic effects. 
To validate our theoretical template, in this section we use \textit{only the catalog with all systematics included (realistic)}. The results of the analysis with the Outer Rim QSO mocks without systematic effects (reference) are presented in App.~\ref{app:ORnosys}.

We fit the mean of 100 realisation of the Outer Rim mock power spectrum multipoles with $\ell=0,2,4$ using the base one-loop perturbation theory model from Ref.~\cite{Ivanov:2019pdj,Chudaykin:2020aoj,Chudaykin:2020ghx}. The mock spectra are binned with the bin size $\Delta k=0.01\hMpc$. At our level of precision, we integrate the theory predictions in each $k$-bin
in order to account for discreteness effects. 
We have fixed $n_s$ and $\omega_b$ and $M_\nu$ to their fiducial values and vary only $\omega_{cdm}$, $h$ and $\ln(10^{10}A_s)$ in our MCMC analysis, as we will do in the actual data analysis.

In order to get a stronger test of our theory we perform the analysis of Outer Rim mocks with the covariance corresponding to 10 times the cumulative eBOSS volume. To that end, we compute the Gaussian analytic covariance with $10\!\times\! V_{\rm QSO}\approx225 h^{-3}\Gpc^3$~\cite{Chudaykin:2019ock,Wadekar:2019rdu,Wadekar:2020hax,Philcox:2020zyp}. This choice even surpasses the future spectroscopic surveys like DESI~\cite{DESI:2016fyo}, so that it will help us to detect even very small systematic shifts in parameter recovery. In the covariance matrix computation we adopt the constant shot noise of the QSO mocks $P_0^{\rm shot}=4.6\cdot 10^4 h^{-3}\Mpc^3$ \cite{Smith:2020stf}.

\begin{figure}[!thb]
\centering
\includegraphics[width=0.49\textwidth]{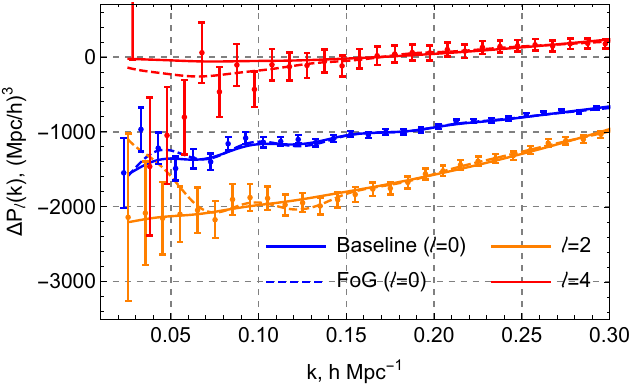}
\caption{Shifts in the power spectrum multipoles induced by the realistic smearing+catastrophic redshift effects applied to the OR mock data (dots). The solid lines show the difference in our best-fit theoretical predictions which includes a constant quadrupole shot noise correction (setting the cosmological parameters to true values). The dashed lines depict results of the standard FoG model. Error bars corresponds to the variance estimated from the Gaussian analytic covariance with the eBOSS volume multiplied by 10.
\label{fig:spect} }
\end{figure}

{\bf Systematic effects.} As a first step, we justify the inclusion of the $P_{\rm shot,2}$. To do so, we perform the following analysis. 
We analyze the Outer Rim data with all injected systematics. 
We fix all cosmological parameters to their true values.
We fit the contaminated Outer Rim mock data with two models: with the  treatment 
of systematics by means of the standard FoG function, and with our theoretical template which includes a constant quadrupole shot noise correction $P_{\rm shot,2}$.
We use the EFT model for the uncontaminated clustering part of the Outer Rim mocks in both cases. Then we plot the difference between the best-fit models for contaminated Outer Rim mocks
and the best-fit model for the "clean" mocks without systematic effects.
The results are shown in Fig.~\ref{fig:spect}.
We see that the standard FoG model fails to capture redshift errors in
the large-scale quadrupole for $k\lesssim 0.05~\hMpc$. In Appendix~\ref{app:FoG}
we show that this model introduces a significant bias in the $\sigma_8$
recovery within the EFT-based full-shape analysis where the redshift 
systematics is modeled
entirely with the standard FoG model.
On the contrary, the EFT model with $P_{\rm shot,2}$ fits the systematic 
effects even on very large scales, and, as we show in Appendix~\ref{app:FoG},
does not induce any bias in the $\sigma_8$ recovery. 

Our best-fit theoretical prediction does not perfectly match the hexadecopole data on large scales, $k<~0.1~\hMpc$. 
In principle, we could adjust the 
model to fit the hexadecapole 
data better 
by introducing 
more nuisance parameters. 
We found, however, that is is not necessary
as this effect does not affect 
cosmological inference from the eBOSS QSO sample. It might be important though for future spectroscopic surveys like DESI and Euclid. 
Note that the short scale systematics, which are relevant for the 
parameter constraints, are absorbed into our nuisance parameters.

As a next step, we also vary cosmological parameters in addition to nuisance parameters.
We assume flat infinite priors
for nuisance parameters in this case, except for 
$b_1,b_2,b_{\mathcal{G}_2}$, for which we use \eqref{biasPr}.
We found that the cosmological parameter recovery is unbiased for $\kmax=0.3\hMpc$. 
From this analysis where we vary cosmology, we find $P_{\rm shot,2}=-1800\pm800$. 
In addition, $P_{\rm shot,2}$ is anti-correlated with the primordial scalar power spectrum amplitude and the mass fluctuation amplitude that broadens significantly their posterior distributions. To lift this degeneracy we choose to employ the conservative prior given by \eqref{pshot2} in what follows. 

We do not impose a tighter prior on $P_{\rm shot,2}$ for the following reasons. First, 
the actual true cosmology of the data is unknown, which introduces an uncertainty in 
the $P_{\rm shot,2}$ measurement from the actual data. Second, the Outer Rim mocks do not contain important observational effects present in the real survey which may also affect the real distribution of $P_{\rm shot,2}$. We explore the impact of
the $P_{\rm shot,2}$ prior on cosmological inference in App. \ref{app:Pshot2}.

{\bf Fingers-of-God modelling.} As a next step we validate our priors on 
EFT 
RSD
counterterms. To do so, we assume the Gaussian prior \eqref{pshot2} along with \eqref{biasPr}. We extract the following leading order counterterm coefficients from the fit to the contaminated OR mocks,
\be
\label{c0c2c4}
\begin{split}
 & c_0(z=1.43)=(30\pm 15)~[h^{-1}\text{Mpc}]^2\,,\\
 & c_2(z=1.43)=(50\pm 30)~[h^{-1}\text{Mpc}]^2\,,\\
 & c_4(z=1.43)=(30\pm 10)~[h^{-1}\text{Mpc}]^2\,.
 \end{split}
\ee
These measurements are consistent with our choice of priors in \eqref{cssPr}. The fingers-of-God effect for the quasars seems to be somewhat stronger when compared to the LRG and ELG samples \cite{Ivanov:2021zmi}~\footnote{Note that QSO measurements correspond to the higher redshift which makes the difference with the LRG/ELG samples more significant.}. But we found that this enhancement is caused by the systematic effects present in the QSO sample rather than the peculiar motion of the quasars, for details see App. \ref{app:ORnosys}. We decided to use the somewhat tighter priors on counterterm parameters as opposed to the previous analyses \cite{Chudaykin:2020aoj,Chudaykin:2020ghx} to mitigate possible prior volume effects.

We found that the next-to-leading order counterterm parameter is tightly constrained by the data
\be
\label{b4}
\tilde c(z=1.43)=(-80\pm 30)~[h^{-1}\text{Mpc}]^4~\,.
\ee
We see that the next-to-leading order FoG effect is weaker for the QSO sample 
than that for the LRG or ELG galaxy samples~\cite{Ivanov:2021zmi}.

As far as stochastic biases are concerned we found that the $P_{\rm shot}$ is consistent the expected shot-noise contribution. The values of $a_2$ are constrained at the level of $1\%$ of $P_0^{\rm shot}$ and are broadly consistent with zero.

\begin{figure}[!thb]
\centering
\includegraphics[width=0.49\textwidth]{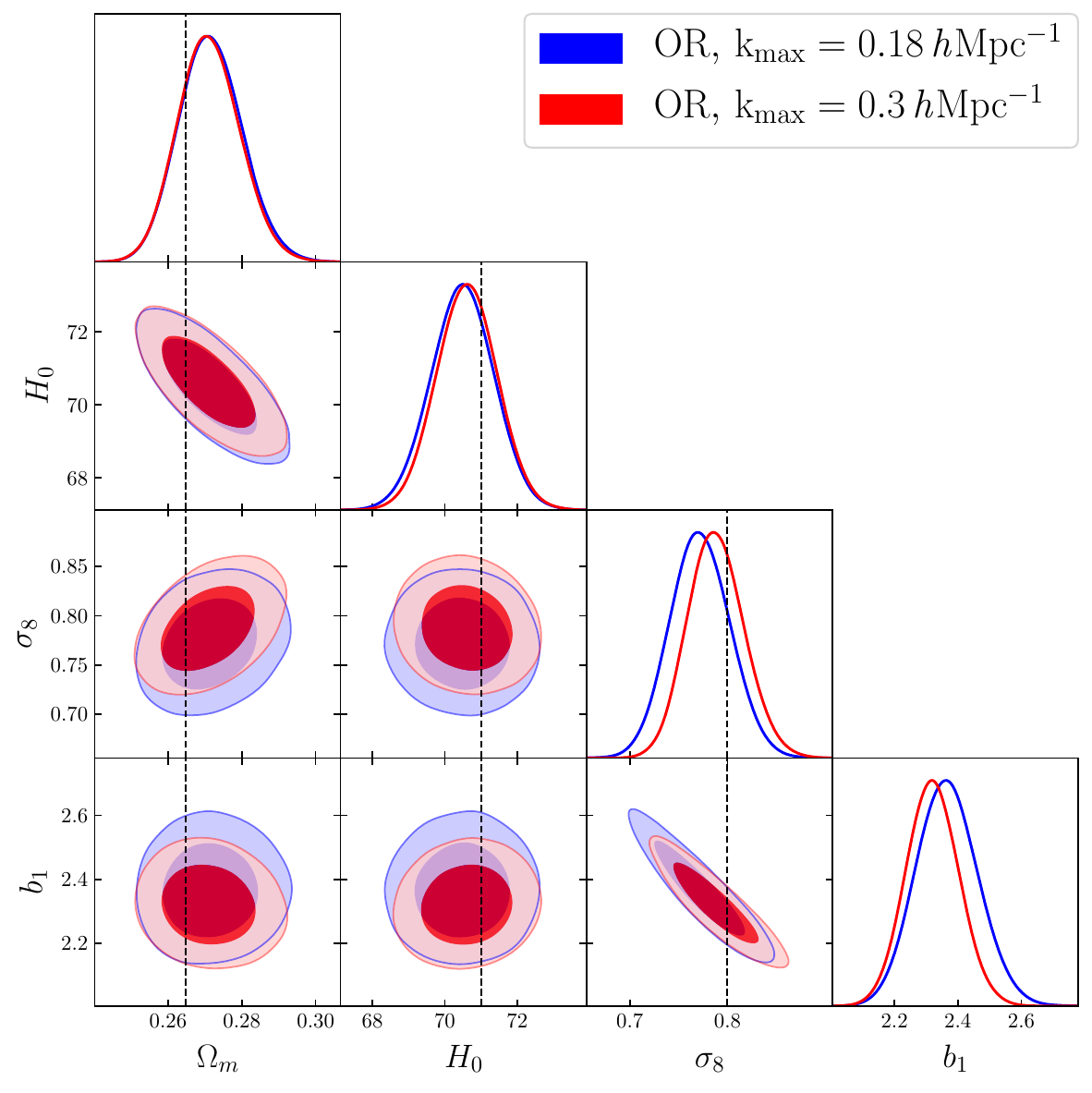}
\caption{Marginalized constraints (68\% and 95\% confidence intervals) on $\Omega_m$, $H_0$, $\sigma_8$ and $b_1$ parameters extracted from the realistic Outer Rim QSO simulation with $10\!\times\!V_{\rm QSO}$ with our baseline model.
\label{fig:ORsys} } 
\end{figure}

{\bf Cosmological constraints.} To ensure that our pipeline recovers true values of cosmological parameters, we fit the mock power spectra for two data cut choices: $\kmax=0.18$ and $\kmax=0.3\hMpc$. The triangular plot for $\Omega_m$, $H_0$ and $\sigma_8$ is shown in Fig. \ref{fig:ORsys}.
The 1d marginalized constraints are listed in Tab. \ref{tab:ORsys}.
\begin{table}[!ht!]
  \begin{tabular}{|c||c|c|c|} \hline
   \diagbox{ { $\kmax $}}{\small Param}  
   &  $\Omega_m$
   & $H_0$
   & $\sigma_8$
      \\ [0.2cm]
\hline
$0.18 \hMpc$   
& $0.2716_{-0.0092}^{+0.0084}$
& $70.48_{-0.89}^{+0.9}$
& $0.7722_{-0.032}^{+0.031}$
\\ 
\hline
$0.3 \hMpc$   
& $0.2711_{-0.009}^{+0.0083}$
& $70.61_{-0.86}^{+0.85}$
& $0.7884_{-0.031}^{+0.028}$
\\ 
\hline
\end{tabular}
\caption{1d marginalized constraints for 
parameters $\Omega_m,H_0,\sigma_8$ extracted 
from the Outer Rim HOD-3 mocks
for several choices of $\kmax$.
The fiducial values are $\Omega_m=0.2648$, $H_0=71$ km/s/Mpc,
$\sigma_8=0.8$.
}
\label{tab:ORsys}
\end{table}

We found that our theoretical pipeline can recover true values of cosmological parameters within $68\%$ CL for all $\kmax$. The maximal bias reaches $0.2\sigma$ of the true statistical error from the eBOSS data at $\kmax=0.3\hMpc$. Remarkably, cosmological measurements depend on a choice of $\kmax$ very weakly. It happens because the power spectrum monopole for the QSO sample is increasingly dominated by shot noise as $\kmax$ increases. It suggests that pushing analysis toward smaller scales leads to very limited cosmological gains. We choose $\kmax=0.3\hMpc$ as a baseline choice following the official eBOSS analysis \cite{Neveux:2020voa}.

All in all, we conclude that our theoretical framework describes the realistic QSO mock data accurately enough in the wavenumber range $k\in[0.02,0.3]\hMpc$.

\subsection{EZ mocks}
\label{sec:mock2}

In this section we test our theoretical pipeline with approximate EZ light-cone mocks. These mocks are based on the effective Zel’dovich approximation and have been used to estimate the impact of observational systematic effects. They capture the light-cone effects, survey geometry, veto masks, radial selections along with various photometric and spectroscopic systematic effects present in the real data. We refer the reader to Ref. \cite{Zhao:2020bib} for the details on the constructing the EZ mock catalogues. 
Each set of mocks consists of 1000 realizations at north and south galactic caps in the redshift range $0.75<z<2.25$. Following \cite{Neveux:2020voa,Hou:2020rse} we model the corresponding spectra as if they were taken from a single snapshot with $z_\eff=1.48$. The choice of the effective redshift should not bias the measurements because the EZ mocks are properly calibrated with the clustering of observed data catalogues measured in corresponding redshift ranges \cite{Zhao:2020bib}. The fiducial EZ mock cosmology is given by
\be
\begin{split}
&h=0.6777\,,\quad \Omega_m=0.307115\,,\quad \omega_b = 0.02214\,,\\
&\omega_{cdm}=0.1189\,,\quad \sigma_8=0.8225\,,\quad n_s=0.9611\,.
\end{split} 
\ee

To disentangle survey geometry effects from the observational systematics, three sets of mocks have been generated. 
The no-systematics mocks include only the survey geometry, light-cone, veto masks, and 
radial selection effects. 
The second kind of EZ mocks includes the shuffling effects used to generate random 
catalogs.
The third and complete EZ mock set include all systematic effects: light-cone, survey geometry, veto masks, radial selections, photometric systematics, shuffling, 
fiber collisions and catastrophic redshift failures. 
These realistic EZ mocks have been used to compute the covariance matrix adopted in the eBOSS analysis \cite{Neveux:2020voa,Smith:2020stf}. The clustering signal in both these mocks was tuned to the one of the eBOSS QSO data by means of the effective bias model \cite{Zhao:2020bib}. In this section we present the analysis of the EZ mocks with all observational systematics applied (realistic, third kind). We test the modeling of separate observational effects in App. \ref{app:EZ}. 

We fit the mean of 1000 EZ mock realisations using the same theoretical model and parameter priors described in Sec. \ref{sec:data4}. We adopt the following data cuts $\kmin=0.02\hMpc$ and $\kmax=0.3\hMpc$ validated earlier in the analysis of the Outer Rim mock data. Since the clustering statistics of the approximate EZ mocks may be inaccurate on small scales \cite{Zhao:2020bib} we perform our analysis with the actual eBOSS covariance estimated from the EZ mocks with all observational systematics applied. As in the previous section, we vary $\omega_{cdm}$, $h$ and $\ln(10^{10}A_s)$ parameters while fixing $n_s$ and $\omega_b$ and $M_\nu$ to their fiducial values. We present our results for $\Omega_m$, $H_0$ and $\sigma_8$.

\begin{figure}[t!]
\begin{center}
\includegraphics[width=0.49\textwidth]{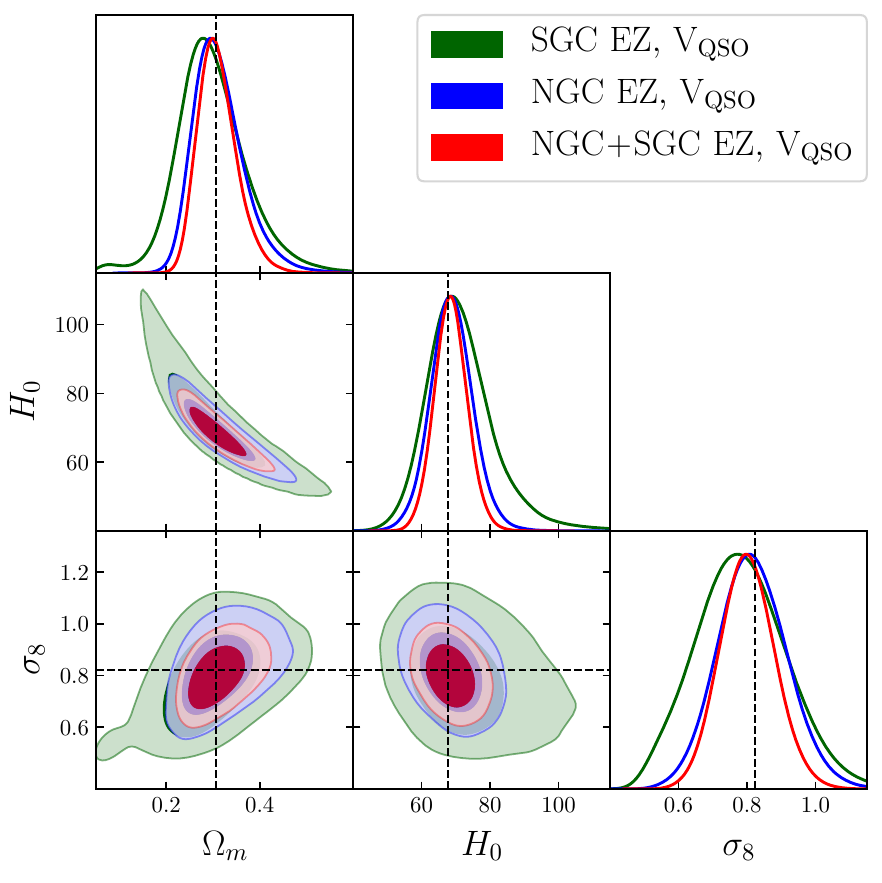}
\end{center}
\caption{
2d and 1d marginalized posteriors from the analysis of the EZ mocks with all observational systematics included and the covariance matching the eBOSS QSO data sample.
\label{fig:EZ}} 
\end{figure}

In Fig. \ref{fig:EZ} and Tab. \ref{tab:EZ} we show our results for the NGC and SGC footprints separately and for their combination.
\begin{table}[!ht!]
  \begin{tabular}{|c||c|c|c|} \hline
   \diagbox{ { Sample}}{\small Param}  
   &  $\Omega_m$
   & $H_0$
   & $\sigma_8$
      \\ [0.2cm]
\hline
SGC  
& $0.2986_{-0.079}^{+0.059}$
& $73.05_{-12.55}^{+5.89}$
& $0.7838_{-0.14}^{+0.13}$
\\ 
\hline
NGC    &  
$0.314_{-0.064}^{+0.043}$
& $68.77_{-6.56}^{+6.}$
& $0.8171_{-0.11}^{+0.1}$
\\ 
\hline 
SGC+NGC   & $0.3106_{-0.049}^{+0.035}$
& $68.55_{-5.09}^{+4.73}$
& $0.802_{-0.085}^{+0.078}$
\\ \hline
\end{tabular}
\caption{1d marginalized intervals of parameters $\Omega_m,H_0$ (in units km/s/Mpc), and $\sigma_8$ extracted 
from the EZ mocks with all observation systematics included.
The fiducial values are $\Omega_m=0.307115$, $H_0=67.77$ km/s/Mpc, $\sigma_8=0.8225$.
}
\label{tab:EZ}
\end{table}
We found that our theoretical pipeline recovers true values of cosmological parameters without any noticeable bias. It justifies our choice of the $P_{\rm shot,2}$ prior \eqref{pshot2}. Another important observation is that the NGC EZ mocks yield more accurate measurements of cosmological parameters. This happens because the effective volume of the NGC QSO sample is two times bigger than the one of the SGC footprint \cite{Smith:2020stf}. It reduces the statistical uncertainty in the parameter recovery. Combining the NGC QSO footprint with the SGC sample one improves the cosmological measurements of $\Omega_m$, $H_0$ and $\sigma_8$ parameters by $\sim20\%$.

\section{Results}
\label{sec:res}

{\bf QSO data.} We present our final results for the QSO sample. We jointly fit the NGC+SGC power spectrum monopole, quadrupole and hexadecopole in the wavenumber range $k\in[0.02,0.3]\hMpc$ (following the official eBOSS analysis~\cite{Neveux:2020voa,Hou:2020rse}). Our results are summarized in Fig. \ref{fig:small} and  Tab.~\ref{tab:main}. We refer the reader to App.~\ref{app:full} for the full parameter constraints.

To put our measurements in context, in Fig. \ref{fig:small} we also show the posteriors of the CMB {\it Planck} 2018 and BOSS FS+BAO+BBN analyses. We found that the cosmological constraints from the eBOSS QSO are fully compatible with the {\it Planck}. The largest deviation is observed in $\sigma_8$ which is $1.6\sigma$ above the {\it Planck} derived value. The agreement between eBOSS QSO and BOSS FS+BAO+BBN is somewhat worse: the estimates of $\sigma_8$ are in a $2.5\sigma$ tension. Note that the $\sigma_8$ measurement from the eBOSS QSO is only $2$ times weaker than the one from the BOSS FS+BAO data analysis. The $\Omega_m$ and $H_0$ measurements exhibit an excellent agreement across all three data sets.

Another important cosmological parameter in the context of the weak-lensing measurements is $S_8\equiv\sigma_8\sqrt{\Omega_m/0.3}$ calculated at the present time. We found from the eBOSS QSO analysis,
\be
S_8=0.976_{-0.12}^{+0.1}
\ee
This value is consistent with the {\it Planck} CMB value $S_8=0.832 \pm 0.013$. It also $1.8-1.9\sigma$ higher than the DES-Y3, $S_8 = 0.776^{+0.017}_{-0.017}$~\cite{DES:2021wwk} and KiDS-1000 $S_8 = 0.759^{+0.024}_{-0.021}$ \cite{KiDS:2020suj} results. 

{\bf Comparison with previous works.}
Let us compare now our results with previous analyses of the eBOSS QSO sample. 
The official eBOSS QSO power spectrum analysis gives
the following values of the Hubble distance, 
the transverse comoving distance, and the linear growth rate 
$D_H(z=1.48)/r_{\rm drag} = 13.52\pm 0.51$
$D_M(z=1.48)/r_{\rm drag} = 30.68\pm 0.90$, and 
$f\sigma_8(z=1.48)=0.476\pm 0.047$.
Our results for the same parameters are
\be 
\begin{split}
 D_H(z=1.48)/r_{\rm drag} &= 12.98_{-0.19}^{+0.21}    \,,\\
 D_M(z=1.48)/r_{\rm drag} &= 30.45\pm 0.78  \,,\\
 f\sigma_8(z=1.48) &= 
0.437_{-0.038}^{+0.036}\,.
\\
\end{split}
\ee
We observe an overall consistency between the two, with the 
errorbars in the full-shape EFT analysis being considerably
smaller, in agreement with arguments given in~\cite{Ivanov:2019pdj,Chudaykin:2020ghx}.

Ref.~\cite{Semenaite:2021aen} presented a joint analysis of 
the eBOSS quasars and BOSS galaxies. This analysis was based 
on a hybrid model involving the RESPRESSO prescription
for the matter power spectrum, full non-linear galaxy bias, 
and a phenomenological non-linear Kaiser model for 
redshift space distortions, supplemented with an additional FoG kernel.
Their results for the eBOSS QSO spectra presented in their Fig. 4 are in good agreement with our measurements.

Finally, Ref.~\cite{Neveux:2022tuk} presented a full-shape analysis 
based on the TNS model and the Gaussian process regression algorithm 
for likelihood sampling. In the setup with fixed $n_s$ and $\omega_b$,
similar to ours, they quote $\Omega_m=0.321\pm 0.016$, $H_0=(65.1\pm 1.9)$ km s$^{-1}$Mpc$^{-1}$,
$\sigma_8 = 1.12\pm 0.10$.
While their results for $\Omega_m$ and $H_0$ are perfectly 
consistent with ours, their value for $\sigma_8$ is more than $1\sigma$ higher than ours, 
which provokes a conclusion of~\cite{Neveux:2022tuk} that the QSO sample alone 
is discrepant with {\it Planck}. The results of our paper show that {\it Planck}
and eBOSS QSO are fully consistent, provided that they are analyzed within the EFT
and with a full MCMC analysis. It would be interesting to understand if the 
insignificant $1\sigma$
discrepancy in $\sigma_8$ between our work and~\cite{Neveux:2022tuk} 
is caused by the theoretical modeling or the likelihood sampling. 
Overall, we conclude that 
our results are nominally consistent with those of~\cite{Neveux:2022tuk}.

{\bf Bias parameters.}
Finally, we present the measurements of the QSO bias parameters. The values of the linear bias extracted from the eBOSS QSO analysis reads 
\be
b^{\rm NGC}_1=2.01_{-0.2}^{+0.17}\,,\quad 
b^{\rm SGC}_1=1.96_{-0.2}^{+0.16} \,,
\ee
Our values are systematically lower than the official eBOSS measurements reported in \cite{Neveux:2020voa,Hou:2020rse}: $b^{\rm NGC}_1=2.39$ and $b^{\rm SGC}_1=2.34$. Note that the latter official values were derived assuming a Planck-like eBOSS fiducial cosmology~\cite{Neveux:2020voa}
with $\sigma_8=0.8$. This value 
is $\approx 1.8~\sigma$ lower
than the value inferred in our 
analysis. 
If one accounts for the fact that the actual best-fit cosmology 
of the eBOSS QSO sample is different from the fiducial one, 
the agreement between us and~\cite{Neveux:2020voa}
will be restored.

For completeness, we provide the measurements of nonlinear bias parameters (see App.~\ref{app:full} for the full cosmological constraints),
\be
\begin{split}
&b^{\rm NGC}_2=-0.76_{-0.93}^{+0.88}\,,\quad 
b^{\rm SGC}_2=-0.28_{-0.92}^{+0.91} \,,\\
&b^{\rm NGC}_{\mathcal{G}_2}=-0.31_{-0.42}^{+0.37}\,,\quad 
b^{\rm SGC}_{\mathcal{G}_2}=-0.18_{-0.46}^{+0.42} \,.
\end{split}
\ee
The $b_2$ measurements are dominated by Gaussian priors \eqref{biasPr}, whereas the quadratic tidal bias $b_{\mathcal{G}_2}$ can be reasonably measured from the QSO sample. This justifies the inclusion of this parameter in a full MCMC analysis.

\section{Discussion}
\label{sec:disc}

We have carried out an EFT-based full-shape analysis of the eBOSS DR16 quasar sample. 
Having fixed the spectral tilt to the Planck $\Lambda$CDM best-fit value 
and the physical baryon density 
to the BBN value, we have obtained $H_0=(66.7\pm 3.2)~$km~s$^{-1}$Mpc$^{-1}$, $\Omega_m=0.32\pm 0.03$, and $\sigma_8=0.945\pm 0.082$ from eBOSS QSO for the minimal $\Lambda$CDM model. 
Our measurements are fully consistent with the predictions of the Planck CMB 
best-fitting $\Lambda$CDM cosmology~\cite{Planck:2018vyg}.

We have confirmed that the main challenge of the full-shape analysis of eBOSS QSO is 
the observation systematics treatment. In particular, the cosmological inference 
of $\sigma_8$ is quite sensitive to redshift smearing effects. We point out that the standard treatment of redshift smearing 
becomes insufficient for the purposes of the EFT-based full-shape analysis, and 
propose a new model that is adequate in the EFT context.
Our phenomenological model is motivated by  simulations, and we have shown that it can
successfully account for
redshift smearing in the eBOSS DR16 sample. 
While this model is adequate given the errorbars
of eBOSS QSO, we believe that future surveys like DESI will require a more
accurate modeling of redshift errors. We hope to develop such a model in future. 
Alternatively, one can try to isolate the line-of-sight modes contaminated by 
redshift systematics, along the lines of the real space reconstruction~\cite{Ivanov:2021fbu}.

Our work can be extended in multiple ways. In this paper we have focused on 
the minimal $\Lambda$CDM model. It would be interesting to extend our analysis to 
non-minimal  
models, for which the spectroscopic clustering data proved to be crucial, 
e.g. spatial curvature~\cite{Chudaykin:2020ghx}, 
dynamical dark energy~\cite{DAmico:2020kxu,Chudaykin:2020ghx}, 
early dark energy~\cite{Ivanov:2020ril,DAmico:2020ods}, massive neutrinos~\cite{Ivanov:2019hqk},
light relics~\cite{Xu:2021rwg}, axion dark matter~\cite{Lague:2021frh}, etc. 

In addition, a natural extension of our analysis would be to include the bispectrum
of the eBOSS QSO sample, and carry out systematic studies along the lines 
of the recent BOSS LRG bispectrum analyses~\cite{Ivanov:2021kcd,Philcox:2022frc,Philcox:2021kcw}. 
This can have significant impact on the 
primordial non-Gaussianity 
constraints, for both local and non-local types~\cite{Cabass:2022wjy,Cabass:2022ymb,DAmico:2022gki}. 

Finally, it would be interesting to estimate the
constraining power of the DESI quasar sample~\cite{DESI:2016fyo} along the lines 
of previous full-shape EFT-based forecasts~\cite{Chudaykin:2019ock,Sailer:2021yzm}. We leave this and other research directions
for future work.

\textit{Note added.} When this work was about to be submitted to arXiv, Ref.~\cite{Simon:2022csv}
with a similar analysis appeared. Our results agree.

\vspace{1cm}
\section*{Acknowledgments}

We are grateful to Richard Neveux for sharing with us the EZ mocks and the integral constraint window functions. We are also grateful to Alex Smith for sharing with us the 
Outer Rim QSO spectra and products. It is a pleasure to thank Stephen Chen
and Nick Kokron
for enlightening discussions.
The work of MMI has been supported by NASA through the NASA Hubble Fellowship grant \#HST-HF2-51483.001-A awarded by the Space Telescope Science Institute, which is operated by the Association of Universities for Research in Astronomy, Incorporated, under NASA contract NAS5-26555.

\appendix 

\section{Full cosmological constraint tables}
\label{app:full}

In this appendix we provide complete constraints on cosmological and nuisance parameters from our analysis of the NGC+SGC eBOSS QSO samples. 
1d marginalized constraints on cosmological parameters are listed in Tab.~\ref{tab:main2}.
\begin{table}[t!]
  \begin{tabular}{|c||c|c|c|} \hline
   Param  
   & mean$\pm\sigma$
   & Param  
   & mean$\pm\sigma$
\\
\hline
$\omega_{cdm}$   
& $0.1184_{-0.011}^{+0.009}$
& $\Omega_m$   
& $0.3205_{-0.038}^{+0.03}$
\\ 
$h$   
& $0.6669_{-0.034}^{+0.031}$
& $\sigma_8$   
& $0.9445_{-0.083}^{+0.081}$
\\ 
$\ln(10^{10}A_s)$  
& $3.375_{-0.18}^{+0.18}$
& $S_8$   
& $0.976_{-0.12}^{+0.1}$
\\  \hline\hline
$b_1^{\rm (NGC)}$  
& $2.01_{-0.2}^{+0.17}$
& $b_1^{\rm (SGC)}$ 
& $1.96_{-0.2}^{+0.16}$
\\ 
$b_2^{\rm (NGC)}$  
& $-0.76_{-0.93}^{+0.88}$
& $b_2^{\rm (SGC)}$ 
& $-0.28_{-0.92}^{+0.91}$
\\ 
$b_{\mathcal{G}_2}^{\rm (NGC)}$  
& $-0.31_{-0.42}^{+0.37}$
& $b_{\mathcal{G}_2}^{\rm (SGC)}$  
& $-0.18_{-0.46}^{+0.42}$
\\ 
$c_{0}^{\rm (NGC)}$  
& $10.8_{-18.88}^{+18.92}$
& $c_{0}^{\rm (SGC)}$  
& $14.4_{-19.54}^{+19.69}$
\\ 
$c_{2}^{\rm (NGC)}$  
& $48.43_{-17.71}^{+18.03}$
& $c_{2}^{\rm (SGC)}$  
& $31.4_{-18.47}^{+18.32}$
\\ 
$c_{4}^{\rm (NGC)}$  
& $21.35_{-16.97}^{+17.03}$
& $c_{4}^{\rm (SGC)}$ 
& $20.03_{-18.04}^{+17.99}$
\\ 
$\tilde c^{\rm (NGC)}$  
& $-23.69_{-70.59}^{+72.02}$
& $\tilde c^{\rm (SGC)}$  
& $-12.7_{-83.01}^{+84.43}$
\\ 
$P_{\rm shot}^{\rm (NGC)}$  
& $0.989_{-0.024}^{+0.025}$
& $P_{\rm shot}^{\rm (SGC)}$  
& $1.002_{-0.022}^{+0.023}$
\\ 
$a_2^{\rm (NGC)}$  
& $0.012_{-0.043}^{+0.041}$
& $a_2^{\rm (SGC)}$ 
& $0.03_{-0.042}^{+0.041}$
\\ 
$P_{\rm shot,2}^{\rm (NGC)}$  
& $-1250_{-687}^{+689}$
& $P_{\rm shot,2}^{\rm (SGC)}$ 
& $-923_{-709}^{+724}$
\\ 
\hline
\end{tabular}
\caption{Mean values and 68\% CL minimum credible
intervals for the cosmological and nuisance parameters of the QSO NGC+SGC sample.
}
\label{tab:main2}
\end{table}

\section{Reference Outer Rim mocks}
\label{app:ORnosys}

In this section we present the full-shape analysis of the ``clean'' Outer Rim QSO mocks (i.e. mocks without injected systematics). We use the usual EFT-based perturbative approach {\it without} a constant term $P_{\rm shot,2}$. We assume the same data cuts as in the baseline analysis: $\kmin=0.02\hMpc$ and $\kmax=0.3\hMpc$.

As a first step, we do not impose any informative priors on nuisance parameters except for the bias coefficients, for which we assume \eqref{biasPr}. We fit the Outer Rim data with the Gaussian
analytic 
covariance corresponding to 10 times the cumulative eBOSS volume. We extract the following leading order counterterm coefficients,
\be
\begin{split}
 & c_0(z=1.43)=(25\pm 15)~[h^{-1}\text{Mpc}]^2\,,\\
 & c_2(z=1.43)=(25\pm 10)~[h^{-1}\text{Mpc}]^2\,,\\
 & c_4(z=1.43)=(15\pm 10)~[h^{-1}\text{Mpc}]^2\,.
 \end{split}
\ee
Interestingly, these measurements 
agree quite well with the results from LRG and ELS samples \cite{Ivanov:2021zmi}.  
Note, however, that the extracted values are considerably smaller than the ones measured from the contaminated Outer Rim QSO mocks, see \eqref{c0c2c4}. Our results
thus indicate that the monopole and quadrupole counterterm coefficients partially absorb the systematic effects present in the
contaminated OR mocks and the
real QSO sample. 
For the next-to-leading order counterterm we get
\be
\tilde c(z=1.43)=(20\pm 30)~[h^{-1}\text{Mpc}]^4~\,.
\ee
This value is consistent with zero, and it is noticeably
smaller than the one extracted from the contaminated OR mocks \eqref{b4}. All in all, our results confirm that couterterm coefficients partially absorb the QSO systematic effects, and that the actual physical 
FoG effect attributed to the
short-scale peculiar velocities 
is weaker for the QSO sample
than the ``effective''
distortions of the line-of-sight 
modes due to redshift systematics.

\begin{figure}[!t]
 \centering
 \includegraphics[width=0.49\textwidth]{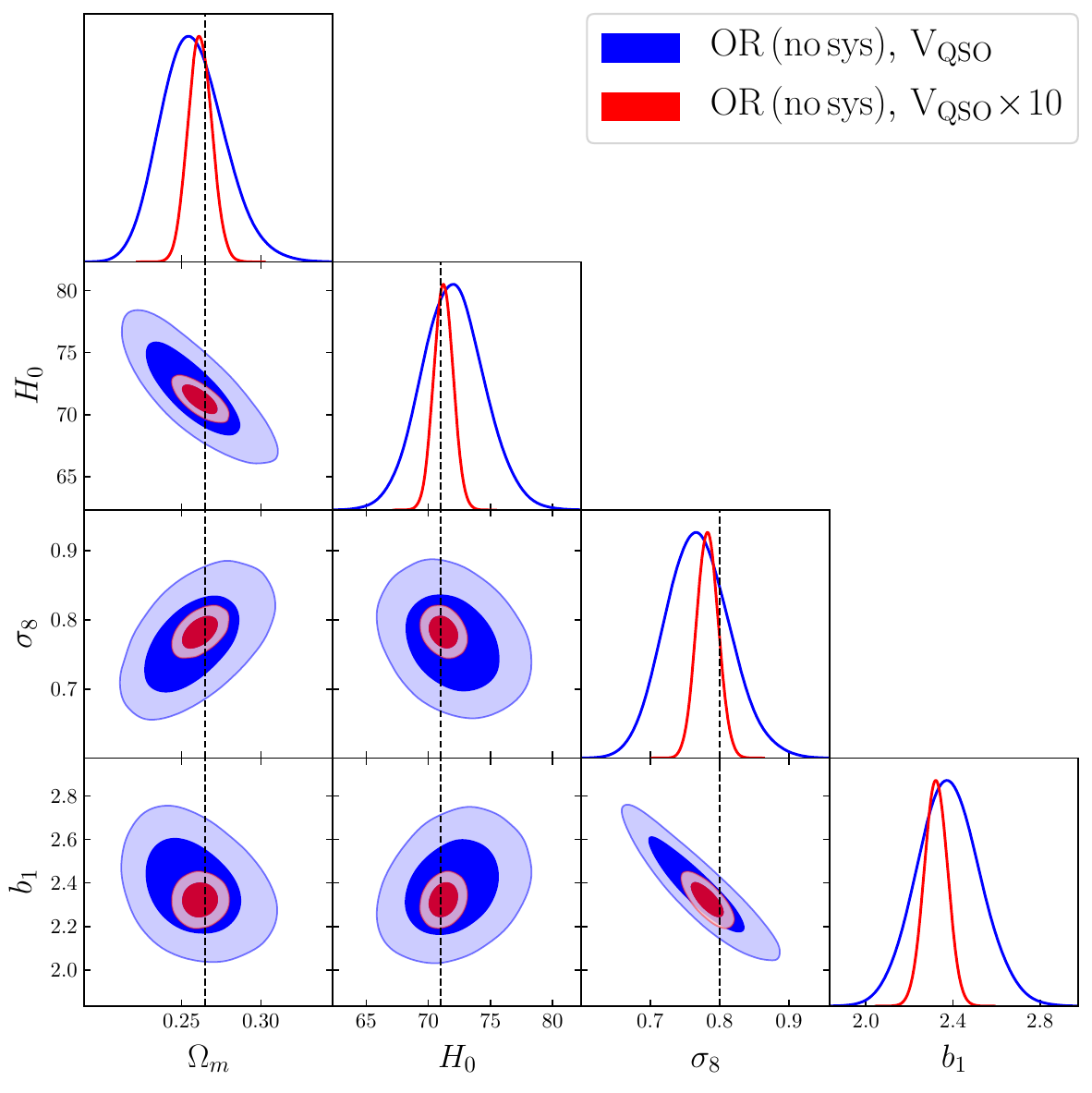}
 \caption{Marginalized constraints (68\% and 95\% confidence intervals) on $\Omega_m$, $H_0$, $\sigma_8$ and $b_1$ parameters extracted from the reference Outer Rim QSO simulation with $V_{\rm QSO}$ (in blue) and $10\!\times\!V_{\rm QSO}$ (in red).
 \label{fig:ORnosys} } 
 \end{figure}

As a second step, we validate our 
full pipeline on the Outer Rim QSO mocks without systematic effects. To do so, we assume Gaussian priors on both the bias parameters \eqref{biasPr} and the counterterm coefficients \eqref{cssPr}. We perform the full-shape analysis with the analytic covariance corresponding to one and 10 times the cumulative eBOSS volume. The posterior distributions of $\Omega_m$, $H_0$, $\sigma_8$ and $b_1$ are shown in Fig. \ref{fig:ORnosys}. 
The 1d marginalized constraints on cosmological parameters are listed in Tab. \ref{tab:ORnosys}.
\begin{table}[!ht!]
  \begin{tabular}{|c||c|c|c|} \hline
   \diagbox{ { Volume}}{\small Param}  
   &  $\Omega_m$
   & $H_0$
   & $\sigma_8$
      \\ [0.2cm]
\hline 
$10\!\times\!V_{\rm QSO}$   
& $0.2616_{-0.0078}^{+0.0075}$
& $71.23_{-0.81}^{+0.8}$
& $0.7824_{-0.016}^{+0.016}$
\\ \hline
$V_{\rm QSO}$  
& $0.2575_{-0.022}^{+0.018}$
& $71.99_{-2.67}^{+2.47}$
& $0.7685_{-0.05}^{+0.045}$
\\  \hline
\end{tabular}
\caption{1d marginalized intervals of parameters $\Omega_m,H_0$ (in units km/s/Mpc), and $\sigma_8$ extracted 
from the OR mocks without systematic effects with one ($V_{\rm QSO}$) and 10 times ($10\!\times\!V_{\rm QSO}$) the cumulative eBOSS volume.
The fiducial values are $\Omega_m=0.2648$, $H_0=71$ km/s/Mpc,
$\sigma_8=0.8$.
}
\label{tab:ORnosys}
\end{table}

In the analysis with 10 times the cumulative eBOSS volume we have found a $1\sigma$ shift in $\sigma_8$. This bias reaches $0.3\sigma$ of the true statistical uncertainty of $\sigma_8$ from the eBOSS QSO sample which we interpret as a 
true theory systematic 
bias in our model. However, when the contaminated Outer Rim mocks are analyzed with a constant term $P_{\rm shot,2}$, this $\sigma_8$ bias diminishes to $0.1\sigma$ of the true statistical error on $\sigma_8$ from the eBOSS data, see Tab.~\ref{tab:ORsys}. This implies 
that in the realistic data analysis both 
the true theory-systematic shifts
and the marginalization projection 
shifts are marginal.

In the analysis with the covariance
that matches 
the actual eBOSS volume,
we see that all the input parameters are recovered within $68\%$ CL. We identified $0.3\sigma$ shifts in $\sigma_8$ and $H_0$, and a $0.2\sigma$ bias in the $\Omega_m$ recovery when compared to the result with the 10 times larger volume. In App. \ref{app:priorV} we show that these shifts are induced entirely by the marginalization projection effects. Conservatively, 
we add this uncertainty 
to the total theory systematic bias 
budget. We found that using wider priors on counterterms would
slightly 
enhance the prior volume effects. In particular, using priors from the previous analyses \cite{Chudaykin:2020aoj,Chudaykin:2020ghx} produces a $0.4\sigma$ shift in the $\sigma_8$ recovery. For this reason, we use a slightly more restrictive priors introduced in \eqref{cssPr}. 

We conclude that in the realistic data analysis with the actual eBOSS volume the parameter volume effects represent the dominant part of the apparent shift of cosmological parameters from the true values. The projection effects affect the posteriors of all cosmological parameters at the level of $0.2-0.3\sigma$ in terms 
of the actual errorbars of the eBOSS QSO sample.

\section{QSO systematics}
\label{app:FoG}

We test the accuracy of the theoretical template for non-linear 
redshift space distortions
and redshift systematics 
used in the official eBOSS analysis \cite{Neveux:2020voa,Smith:2020stf}. 
They model non-linear corrections due to Fingers-of-God or virial motions by a phenomenological FoG prefactor.
To match the official eBOSS analysis we model the QSO power spectrum as follows
\be
\label{fog}
P^{\rm FOG}(k,\mu)=D(k,\mu) P_{\rm SPT}(k,\mu)\,,
\ee
where $P_{\rm SPT}(k,\mu)$ is the one-loop power spectrum for bias tracers and $D(k,\mu)$ is a two-parameter damping function, which comprises a Gaussian and a Lorentzian-like terms, \cite{Hou:2018yny}
\be
\label{fog2}
D(k,\mu)=\frac{1}{\sqrt{1+(k\mu a_{\rm vir})^2}}\exp\left[-\frac{(k\mu\sigma_v)^2}{1+(k\mu a_{\rm vir})^2}\right]
\ee
In the last expression $\sigma_v$ is the velocity dispersion and $a_{\rm vir}$ parameterize the kurtosis of the
small scale velocity distribution, for details see \cite{BOSS:2016off,Hou:2018yny}. We do not add an extra exponential term as in \cite{Hou:2018yny} to be consistent with the official eBOSS analysis \cite{Neveux:2020voa}.
 
\begin{figure}[!t]
\centering
\includegraphics[width=0.49\textwidth]{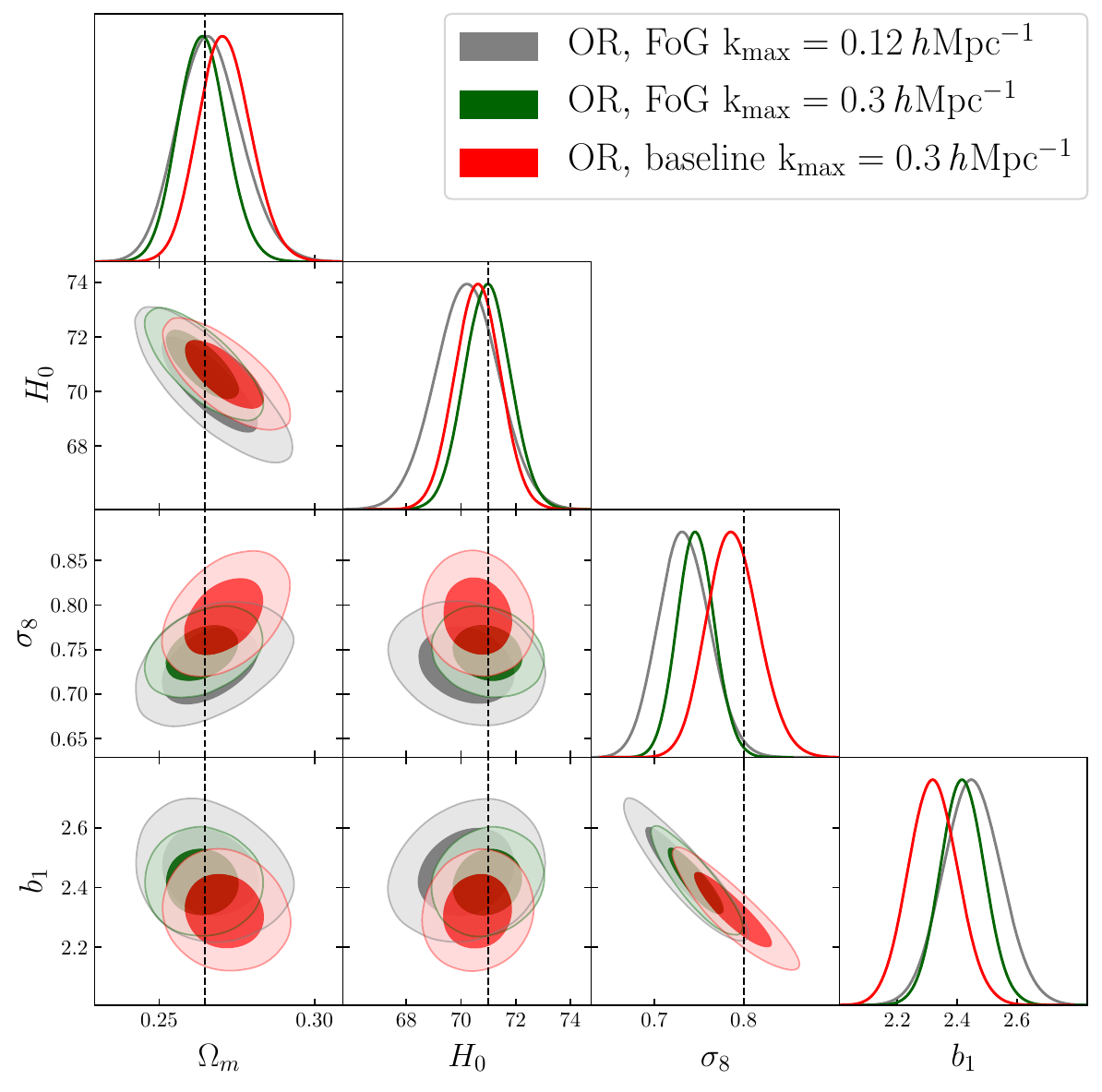}
\caption{Marginalized constraints (68\% and 95\% confidence intervals) on $\Omega_m$, $H_0$, $\sigma_8$ and $b_1$ parameters extracted from the realistic Outer Rim QSO simulation with $10\!\times\!V_{\rm QSO}$ using the standard FoG model and our baseline analysis pipeline.
\label{fig:ORapp} } 
\end{figure}

We estimate now the performance of this standard FoG model. To do so, we fit the contaminated Outer Rim mock data for two data cut choices: $\kmax=0.12$ and $\kmax=0.3\hMpc$. We perform our analysis with the Gaussian covariance corresponding to 10 times the cumulative eBOSS volume. We vary all cosmological and nuisance parameter along the lines of Sec. \ref{sec:data4} assuming \eqref{biasPr} and \eqref{c0c2c4}. 
The triangular plot for $\Omega_m$, $H_0$, $\sigma_8$ and $b_1$ is shown in Fig. \ref{fig:ORapp}.
The 1d marginalized constraints on cosmological parameters are listed in Tab. \ref{tab:ORfog}.
\begin{table}[!ht!]
  \begin{tabular}{|c||c|c|c|} \hline
   \diagbox{ { $\kmax$}}{\small Param}  
   &  $\Omega_m$
   & $H_0$
   & $\sigma_8$
      \\ [0.2cm]
\hline
$0.12\hMpc$  
& $0.2664_{-0.011}^{+0.01}$
& $70.23_{-1.2}^{+1.19}$
& $0.7339_{-0.03}^{+0.028}$
\\ 
\hline
$0.3\hMpc$  
& $0.2639_{-0.008}^{+0.008}$
& $70.97_{-0.88}^{+0.84}$
& $0.7468_{-0.022}^{+0.021}$
\\ 
\hline 
\end{tabular}
\caption{1d marginalized intervals of parameters $\Omega_m,H_0$ (in units km/s/Mpc), and $\sigma_8$ extracted from the OR mocks with all systematics included within the standard FoG theoretical framework.
The fiducial values are $\Omega_m=0.2648$, $H_0=71$ km/s/Mpc, $\sigma_8=0.8$.
}
\label{tab:ORfog}
\end{table}

We found that 
the standard FoG model fails to reproduce the mock data the even on large scales. It leads a $2.3\sigma$ and $2.5\sigma$ shift in the $\sigma_8$ recovery for $\kmax=0.12\hMpc$ and $\kmax=0.3\hMpc$, respectively. We conclude that the standard FoG modeling introduces a significant bias within the EFT-based full-shape analysis. This bias originates from the large-scale quadrupole moment in the range $k\lesssim0.05\hMpc$ according to Fig. \ref{fig:spect}. To restore an agreement with the quadrupole data on large scales~\footnote{Note that the small scale quadrupole data is reproduced by fitting the FoG parameters $\sigma_v$ and $a_{\rm vir}$.}, the mass fluctuation amplitude decreases that explains the systematic shift in $\sigma_8$ found in Tab. \ref{tab:ORfog}. As the combination $\sigma_8^2b_1^2$ is fixed by the amplitude of the monopole power spectrum, the standard FoG modeling also systematically shifts the linear bias by amount $\Delta b_1=+0.1$, see Fig. \ref{fig:ORapp}.

\section{Modeling observational effects}
\label{app:EZ}

We validate the modeling of separate observational effects applied to the light-cone EZ mocks. 
 
\begin{figure}[ht!]
\begin{center}
\includegraphics[width=0.49\textwidth]{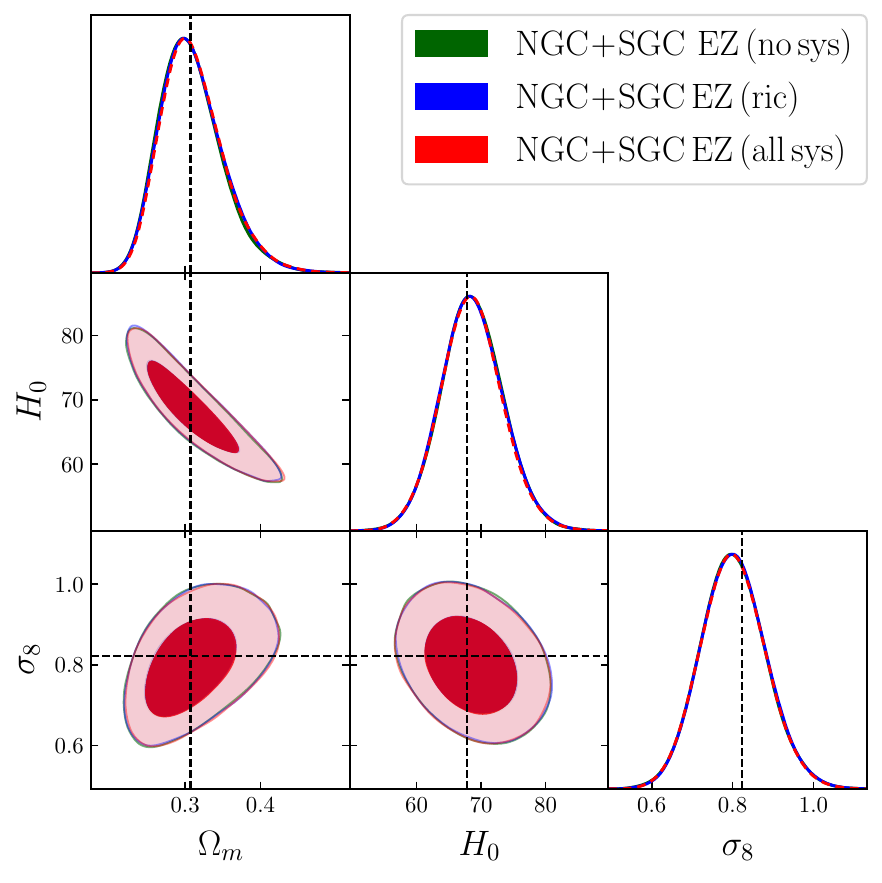}
\end{center}
\caption{
2d and 1d marginalized posteriors
from the analysis of the EZ mocks without observation systematic effects (in green), with the radial integral constraint (in blue) and with all systematics applied (in red).
\label{fig:ezref} } 
\end{figure}

The posterior distributions for the cosmological parameters extracted from the EZ mock without observational effect,  with the radial integral constraint and with all systematics included are shown in Fig.~\ref{fig:ezref}.
We find almost identical cosmological constraints for all scenarios after appropriate theoretical modeling of the
relevant observational effects. 

\section{Prior volume effects}
\label{app:priorV}

We examine the marginalization projection effects present in the posterior distribution of cosmological parameters. In the analysis of the reference Outer Rim QSO mocks (see App.~\ref{app:ORnosys}) we identify $0.2-0.3\sigma$ shifts in parameter recovery. Here we show that this difference is entirely explained by prior volume effects.

To estimate prior volume effects we generate a synthetic data vector without any statistical scatter using the fitting pipeline. As a first step, we produce the best-fit theory curve from the analysis of the reference Outer Rim QSO mocks with $\kmax=0.3\hMpc$ and the Gaussian covariance with volume $10\!\times\! V_{\rm QSO}$~\footnote{Note that the best-fit curve does not depend on the effective volume of the covariance. We assume $10\!\times\! V_{\rm QSO}$ to improve the convergence of MCMC chains.}. 
As a next step, we fit this synthetic data vector with the same theoretical model as if this were actual data. Since the mock data vector is generated with the 
same theory pipeline, and there is no statistical noise in the data, 
it is guaranteed that the true input parameters would precisely minimize 
the mock data likelihood. Hence, any shift of the posterior away 
from the input parameters are entirely due to 
Bayesian prior volume (marginalization) effects.

\begin{figure}[!t]
\centering
\includegraphics[width=0.49\textwidth]{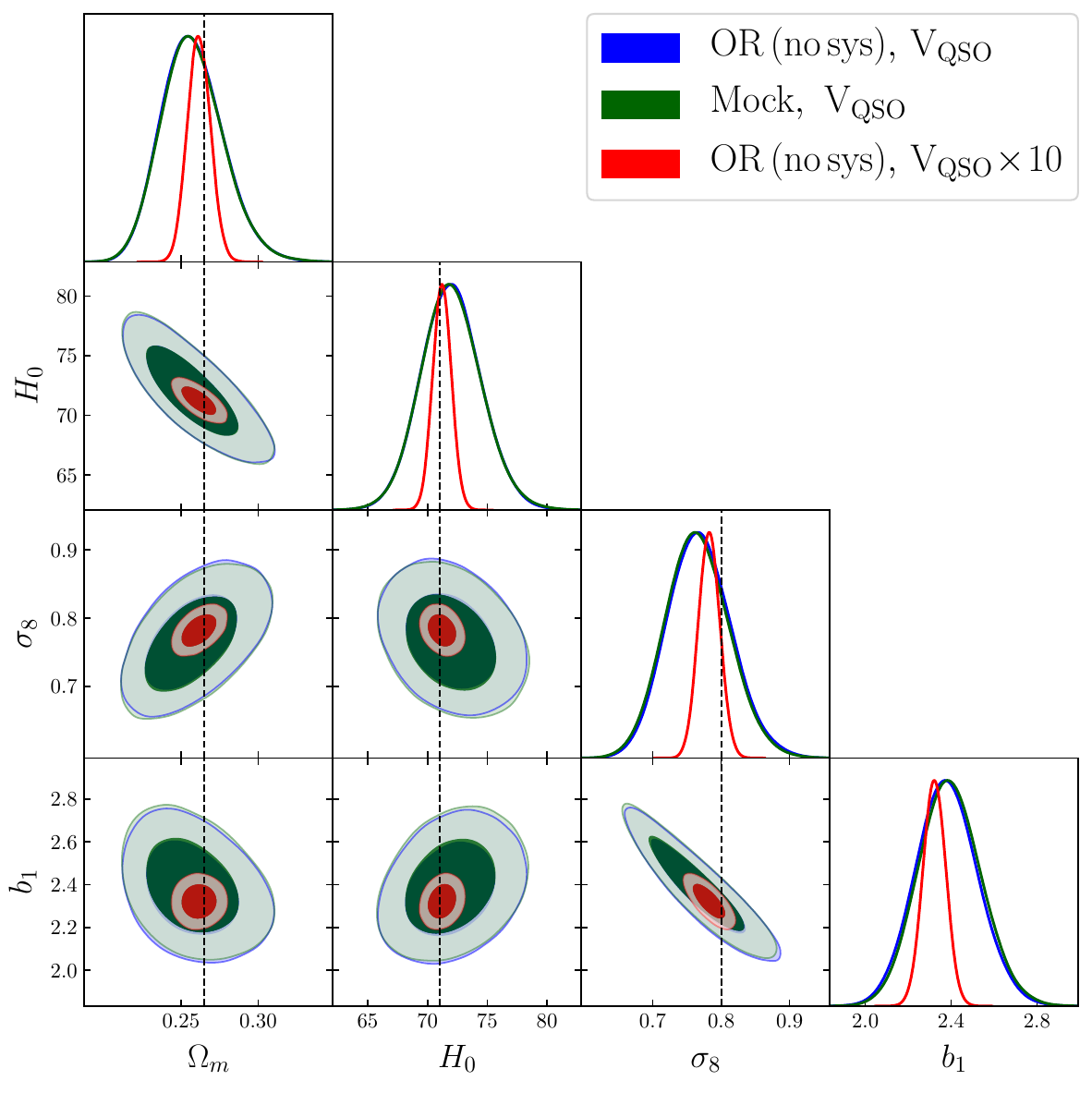}
\caption{Marginalized constraints (68\% and 95\% confidence intervals) on $\Omega_m$, $H_0$, $\sigma_8$ and $b_1$ extracted from the reference Outer Rim QSO simulation with $V_{\rm QSO}$ (in blue) and $10\!\times\!V_{\rm QSO}$ (in red), and from the noiseless mock data (in green) computed with our pipeline for the bestfit cosmology at $\kmax=0.3\hMpc$.
\label{fig:priorV0p5} } 
\end{figure}

First, we perform the analysis of our noiseless mock data with the covariance corresponding to the actual eBOSS volume. In the absence of the projection effects, our model is expected to accurately recover the input cosmology computed with our theoretical pipeline at the same $\kmax=0.3\hMpc$. 
However, when we actually 
fit this mock data, we find that the means of the cosmological parameters are shifted as shown in Fig.~\ref{fig:priorV0p5}. 
Remarkably, the marginalized constraints closely match those obtained from the actual Outer Rim QSO mocks with the covariance whose effective volume corresponds to $V_{\rm QSO}$.
As the actual Outer Rim mocks and noiseless mock data give identical results, 
the shifts in the parameter posteriors should be driven entirely by prior volume effects.
We conclude that the Bayesian marginalization shifts do not exceed $0.3\sigma$.

\begin{figure}[t!]
\centering
\includegraphics[width=0.49\textwidth]{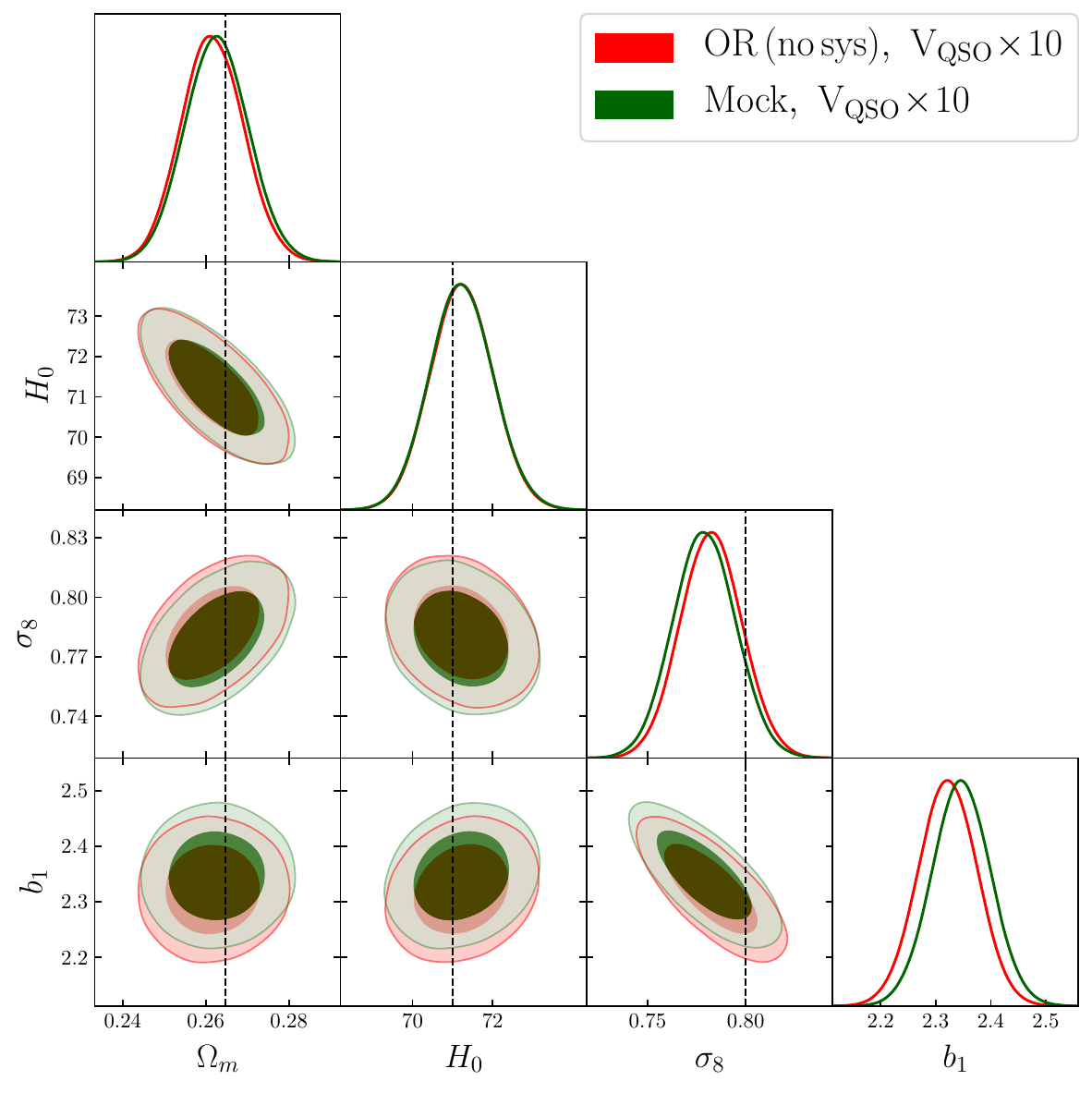}
\caption{Marginalized constraints (68\% and 95\% confidence intervals) on $\Omega_m$, $H_0$, $\sigma_8$ and $b_1$ extracted from the reference Outer Rim QSO simulation (in red), and from the noiseless mock data (in green) computed with our pipeline for the bestfit cosmology at $\kmax=0.3\hMpc$. The covariance corresponds to 10 times the actual eBOSS volume in both cases.
\label{fig:priorV5} } 
\end{figure}

As a next step, we estimate the prior volume effects with 10 times the cumulative eBOSS volume. The resulting posterior distributions are shown in Fig. \ref{fig:priorV5}.
We found that the projections effects are marginal in this case. Indeed, in the case of the larger sample volume statistical errors shrink, so that the marginalization projection effects become smaller.
This is important to keep in mind 
when discussing prior volume effects: 
they disappear once we have more data
and hence the constraints become less sensitive
to the priors. In can be contrasted 
with the actual theory systematic 
error which does not 
depend on the prior volume or 
the quality of the data. 

\section{$P_{\rm shot,2}$ prior}
\label{app:Pshot2}

We examine the sensitivity of the cosmological constraints to the choice of $P_{\rm shot,2}$ prior. To that end, we perform the analyses of the QSO mocks with the more restrictive Gaussian prior $P_{\rm shot,2}=-1800\pm800$ extracted from the OR mock data (see Sec. \ref{sec:mock1}).

First, we present analysis of the OR mocks with all systematics applied. To that end, we fit the contaminated Outer Rim data with $\kmax=0.3\hMpc$ using the analytic covariance corresponding to 10 times the cumulative eBOSS volume. Following the analysis in Sec. \ref{sec:mock1} we vary all cosmological and nuisance parameter assuming \eqref{biasPr} and \eqref{c0c2c4} but with the more stringent prior $P_{\rm shot,2}=-1800\pm800$. 1d marginalized constraints on cosmological parameters are listed in Tab. \ref{tab:Pshot2OR}. For reference, we also show the cosmological constraints with the baseline choice of prior \eqref{pshot2}.
\begin{table}[!t]
  \begin{tabular}{|c||c|c|c|} \hline
   \diagbox{ { $\kmax $}}{\small Param}  
   &  $\Omega_m$
   & $H_0$
   & $\sigma_8$
      \\ [0.2cm]
\hline
New prior 
& $0.2715_{-0.0091}^{+0.0083}$
& $70.62_{-0.86}^{+0.85}$
& $0.7960_{-0.031}^{+0.028}$
\\ 
\hline
Baseline  
& $0.2711_{-0.009}^{+0.0083}$
& $70.61_{-0.86}^{+0.85}$
& $0.7884_{-0.031}^{+0.028}$
\\ 
\hline
\end{tabular}
\caption{1d marginalized intervals of parameters $\Omega_m$, $H_0$ (in units km/s/Mpc), and $\sigma_8$ extracted 
from the realistic Outer Rim QSO simulation using the baseline prior \eqref{pshot2} (Baseline) and $P_{\rm shot,2}=-1800\pm800$ (New prior). The measurements are performed with 10 times the cumulative eBOSS volume.
The fiducial values are $\Omega_m=0.2648$, $H_0=71$ km/s/Mpc,
$\sigma_8=0.8$.
}
\label{tab:Pshot2OR}
\end{table}

We found that the measurements with the more restricted prior on $P_{\rm shot,2}$ are fully consistent with the baseline analysis. In particular, the constraints on $\Omega_m$ and $H_0$ remain essentially the same. We identified a $0.2\sigma$ shift in the $\sigma_8$ recovery in terms of the statistical uncertainty. Note that in the realistic data analysis with the actual eBOSS volume this shifts is negligible.

Next, we perform the analysis of the EZ mocks with all systematic applied. Following the analysis in Sec. \ref{sec:mock2} we vary all cosmological and nuisance parameters with the covariance matching the eBOSS QSO data sample. We present the combined results for the NGC and SGC galactic caps.
The 1d marginalized constraints are listed in Tab. \ref{tab:Pshot2EZ}.
\begin{table}[!ht!]
  \begin{tabular}{|c||c|c|c|} \hline
   \diagbox{ { $\kmax $}}{\small Param}  
   &  $\Omega_m$
   & $H_0$
   & $\sigma_8$
      \\ [0.2cm]
\hline
New prior   & $0.3018_{-0.046}^{+0.034}$
& $69.43_{-5.15}^{+4.69}$
& $0.822_{-0.087}^{+0.079}$
\\ \hline
Baseline   & $0.3106_{-0.049}^{+0.035}$
& $68.55_{-5.09}^{+4.73}$
& $0.802_{-0.085}^{+0.078}$
\\ \hline
\end{tabular}
\caption{1d marginalized constraints for 
parameters $\Omega_m$, $H_0$, $\sigma_8$ extracted from the realistic EZ mocks (NGC+SGC) using the baseline prior \eqref{pshot2} (Baseline) and $P_{\rm shot,2}=-1800\pm800$ (New prior). The measurements are performed with the covariance that matches 
the actual eBOSS volume.
The fiducial values are $\Omega_m=0.307115$, $H_0=67.77$ km/s/Mpc, $\sigma_8=0.8225$.
}
\label{tab:Pshot2EZ}
\end{table}

We found that the cosmological measurements with the more restrictive prior on $P_{\rm shot,2}$ barely change. The difference with respect to the baseline analysis for all cosmological parameter is within $0.2\sigma$ of the true statistical uncertainty. 

We conclude that the baseline analysis is robust against the choice of the $P_{\rm shot,2}$ prior.

\newpage 

\bibliography{short.bib}

\end{document}